\documentclass[a4paper,11pt]{article}
\pdfoutput=1 

\usepackage{jcappub} 

\usepackage[T1]{fontenc} 

\usepackage{blindtext}
\usepackage{hyperref}
\usepackage{amsmath,amssymb}
\usepackage{float}
\usepackage{microtype}
\usepackage{ragged2e}  

\usepackage{graphicx}
\usepackage{bm}
\usepackage{latexsym}
\usepackage{epsfig}
\usepackage{psfrag}
\usepackage[dvipsnames]{xcolor}
\usepackage{subfigure}
\usepackage{subcaption}
\usepackage{graphicx}
\usepackage{multirow}
\usepackage{amssymb}
\usepackage{amsmath}
\usepackage{bm}
\usepackage{latexsym}
\usepackage{epsfig}
\usepackage{psfrag}
\usepackage[normalem]{ulem}
\usepackage{textcomp}
\usepackage{color}
\usepackage{pstricks}
\usepackage[utf8]{inputenc}
\usepackage{comment}
\usepackage{hyperref}
\usepackage[mathlines]{lineno}
\usepackage{hyperref}
\usepackage[all]{hypcap}
\usepackage{subfigure}
\usepackage{caption}
\usepackage{subcaption}

\hypersetup{  
    colorlinks=true,
    linkcolor=blue,
    filecolor=blue,      
    urlcolor=blue,
    citecolor=blue,
    }

\title{Primordial magnetic fields in the light of upcoming post-EoR Lyman-$\alpha$ and 21-cm observations}

\author[a,1]{Arko Bhaumik, \note{Corresponding author.}}
\author[a]{Sourav Pal,}
\author[a]{Supratik Pal}

\affiliation[a]{Physics and Applied Mathematics Unit, Indian Statistical Institute, \\ 203, B.T. Road, Kolkata 700108, India}

\emailAdd{arkobhaumik12@gmail.com}
\emailAdd{soupal1729@gmail.com}
\emailAdd{supratik@isical.ac.in}

\abstract{The Lorentz force exerted by a primordial magnetic field (PMF) on the coupled baryon-dark matter system may enhance total matter power at small scales after recombination. In the post-reionization (post-EoR) era, a weakly scale-dependent PMF of sub-nG strength is thus expected to influence the Lyman-$\alpha$ (Ly$\alpha$) power spectrum, the 21 cm power spectrum, and the Ly$\alpha$-21 cm cross-spectrum at scales $k\gtrsim 1\:h/\textrm{Mpc}$. We investigate the prospects of constraining the PMF sector via these three cosmological observables, 
by employing SNR estimation and Fisher forecast on the PMF amplitude $B_0$ and spectral index $n_{\rm B}$, for a next-generation DESI-like spectroscopic survey and two upcoming 21 cm facilities, namely SKA1-Mid and PUMA. Our results indicate the possibility of constraining both PMF parameters with $\lesssim10\%$ relative errors through the uncontaminated 21 cm auto-spectrum as well as the Ly$\alpha$-21 cm cross-spectrum probed with the DESI-like+SKA1-Mid combination. Indicatively, the Ly$\alpha$-21 cm cross-correlation via DESI-like+SKA1-Mid is predicted to constrain a fiducial scenario $B_0=0.8$ nG and $n_{\rm B}=-2.9$ with $1\sigma$ errors $\Delta B_0\approx 0.07$ nG and $\Delta n_{\rm B}\approx0.02$. The DESI-like+PUMA setup is predicted to fare relatively worse due to its restriction to larger scales, resulting in comparatively one order of magnitude relaxed error bounds for similar fiducials. Since the Ly$\alpha$-21 cm cross-signal is expected to be largely insensitive to foreground contamination (unlike the 21 cm auto-spectrum), it may serve as an optimal foreground-immune post-EoR probe to constrain a weakly scale-dependent sub-nG PMF via future DESI-like+SKA1-Mid observations.}

\begin{document}
\maketitle
\flushbottom

\section{Introduction}
\label{sec:intro}

Primordial magnetic fields (PMFs) present in intergalactic voids on length scales of $\mathcal{O}$(Mpc) have a wide range of allowed magnitude $10^{-16}\:\textrm{G}\lesssim B_0\lesssim 10^{-9}\:\textrm{G}$ based on current data. The lower bound is based on observations of $\gamma$-ray cascades associated with distant blazars \cite{2010Sci...328...73N,2011ApJ...727L...4D,Tiede:2017aql}, whereas the upper bound is furnished by Faraday rotation measures of extragalactic sources \cite{Blasi:1999hu,2010ApJ...723..476A,Vernstrom:2019gjr,OSullivan:2020pll} and Cosmic Microwave Background (CMB) constraints \cite{Barrow:1997mj,2009PhRvL.103h1303S,Planck:2015zrl,POLARBEAR:2015ktq,Zucca:2016iur,Sutton:2017jgr}.\footnote{More recently, it has been shown that certain models of PMF-induced baryon clumping during the pre-recombination era may further tighten the upper bound to $\sim10$ pG on the basis of current CMB data \cite{Jedamzik:2020krr,Galli:2021mxk,Jedamzik:2025cax}.} At small cosmological scales, a nearly scale-invariant stochastic Gaussian-distributed PMF of non-helical nature is theorized to enhance structure formation via an effective Lorentz force term which induces growth of density perturbations, leading to a localized bump in the total matter power spectrum typically at comoving scales $k\gtrsim1\:h/\textrm{Mpc}$ for $B_0\lesssim1$ nG \cite{2012PhRvD..86d3510S,2012ApJ...748...27P,Kunze:2022mlr,Adi:2023doe,Ralegankar:2024ekl,Ralegankar:2024arh}. Such PMF-boosted matter power may significantly affect early star formation history \cite{2020A&A...643A..54S,Koh:2021gjt,Zhang:2024yph}, thus leaving detectable imprints on 21-cm observables during the Cosmic Dawn and the Epoch of Reionization (EoR) \cite{Pandey:2014vga,Adi:2023qdf,Cruz:2023rmo,Bhaumik:2024efz}. On the other hand, in the post-EoR Universe at redshift $z\lesssim6$, the Lyman-$\alpha$ (Ly$\alpha$) forest, which consists of absorption features in the background quasar spectra, may serve as a crucial cosmological observable for constraining PMFs based on small scale matter clustering \cite{Kahniashvili:2012dy,2013ApJ...762...15P,Chongchitnan:2013vpa}. However, the fact that the Ly$\alpha$ forest arises mainly from the intergalactic medium (IGM) away from the galactic cores mitigates contamination from astrophysically generated magnetic fields. This, in turn, makes the Ly$\alpha$ one-dimensional (1D) flux power spectrum a particularly clean probe of PMFs at small scales, with current data \cite{Boera:2018vzq} already indicating a sub-nG upper bound on $B_0$ \cite{Pavicevic:2025gqi}. 

Besides the Ly$\alpha$ sector, another crucial observation to probe small scale physics in the late Universe is furnished by 21-cm intensity mapping (IM). In the post-EoR Universe, neutral hydrogen (HI) is mostly concentrated within galaxies, whose spatial distribution traces the underlying dark matter (DM) density field. Hence, on sufficiently large linear scales $k\lesssim0.1\:h/\textrm{Mpc}$, 21-cm fluctuations trace dark matter density perturbations, which makes the 21-cm power spectrum a biased tracer of the total matter power spectrum at $2\lesssim z\lesssim6$ \cite{2010MNRAS.407..567B,2012MNRAS.421.3570G}. At smaller (\emph{i.e.} nonlinear) scales $k\gtrsim0.5\:h/\textrm{Mpc}$, the HI bias becomes a function of both scale and redshift, which may be approximated analytically at the leading order from the concordant $\Lambda$CDM model-based hydrodynamical simulations \cite{Sarkar:2016lvb}. Thus, any small scale enhancement of the matter power due to the presence of a PMF should also be reflected in the post-EoR 21-cm power spectrum at the aforesaid nonlinear scales, albeit showing up with a modified scale dependence. 

Besides the auto-correlation functions of the Ly$\alpha$ forest and the HI 21 cm IM, a third important window into the physics of the post-EoR Universe is offered by the Ly$\alpha$-21 cm cross-correlation signal \cite{2011MNRAS.410.1130G,Sarkar:2015jta,Carucci:2016yzq}. Although the Ly$\alpha$ flux and the 21 cm fluctuations emerge from different physical processes, their cross-power spectrum is expected to trace the total matter power spectrum with a nonlinear bias at small scales, thereby serving as a robust probe of any potential deviation of matter power from its baseline model at such scales. Recent observational data from the Canadian Hydrogen Intensity Mapping Experiment (CHIME), combined with measurements of the 3D Ly$\alpha$ forest power spectrum from the extended Baryon Oscillation Spectroscopic Survey (eBOSS) \cite{deBelsunce:2024knf}, have indicated the detection of 21 cm emission at an average redshift of $\bar{z}\sim2.3$ based on the Ly$\alpha$-21 cm cross-correlation function \cite{CHIME:2023til}. In the near future, this observable could prove to be a particularly promising probe for constraining models of non-cold dark matter \cite{Sarkar:2019yea} and late-time dynamical dark energy \cite{Dash:2020yuq}, testing modified gravity models \cite{Dash:2020yfq}, and studying reionization relics in the post-EoR intergalactic medium (IGM) \cite{Montero-Camacho:2024xvf}.

Past and presently ongoing surveys, \emph{e.g.}, the Dark Energy Spectroscopic Instrument (DESI) \cite{DESI:2016fyo,DESI:2022xcl,DESI:2024lzq,DESI:2025zpo}, the UV-Visual Echelle Spectrograph (UVES) of the Very Large Telescope (VLT) \cite{2000SPIE.4008..534D,2003MNRAS.346..209L,Kim:2003qt,2005MNRAS.362..549S}, and the High Resolution Echelle Spectrometer (HIRES) of the Keck telescope \cite{1992ESOC...40..223V,Lu:1996sn,Day:2019joh}, have already provided us with estimates of the Ly$\alpha$ 1D flux power on small cosmological scales in the post-reionization era. In the coming years, the possibility of precisely probing the post-EoR Ly$\alpha$ flux power spectrum across the redshift range $2<z<6$ is well within reach with a Stage-V DESI-like spectroscopic survey, that may be able to optimistically detect up to $70-100$ Ly$\alpha$ quasi-stellar object (QSO) samples at $z\gtrsim2$ per unit redshift per unit magnitude per degree\textsuperscript{2} \cite{Chaussidon:2022pqg}. On the other hand, the Ly$\alpha$-21 cm cross-power spectrum would be measurable through the functional synergy of such DESI-like spectroscopic surveys and the Square Kilometre Array Phase 1 Mid-Frequency (SKA1-Mid) \cite{2015aska.confE..86N,SKA:2018ckk,Weltman:2018zrl}, as well as with proposed successor designs of SKA1-Mid like the Packed Ultra-wideband Mapping Array (PUMA) \cite{CosmicVisions21cm:2018rfq,PUMA:2019jwd,Castorina:2020zhz}. To constrain a PMF of sub-nG strength via its impact on the 21 cm sector, the SKA1-Mid facility would be particularly suitable due to its long effective baseline $\sim 150$ km of total antennae distribution, which should allow it to probe small comoving scales $k\gtrsim1.0\:h/\textrm{Mpc}$. With this in mind, we perform a detailed forecast analysis to shed light on the potential synergy between a next-generation DESI-like spectroscopic survey on one hand, and the upcoming radio telescope projects SKA1-Mid and PUMA on the other hand, thereby assessing their joint ability to constrain the PMF sector via observations of the Ly$\alpha$ forest power spectrum, the 21 cm power spectrum, and the Ly$\alpha$-21 cm cross-power spectrum in the post-EoR era. Our analysis explicitly compares among the efficacy of these three late time cosmological observables in placing constraints on the parameters of a stochastic non-helical PMF over the coming decade. 

The paper is organized as follows. In Sec. \ref{sec:pmf_matt}, we have briefly reviewed the impact of a non-helical stochastic PMF with spectral index $-3<n_{\rm B}<-1.5$ on matter power at small scales. In Sec. \ref{sec:pmf_ly_21}, the theoretical modeling of the Ly$\alpha$ flux and 21 cm power spectra at small cosmological scales has been summarized, alongside a description of the noise models for the proposed spectroscopic DESI-like instrument and the radio facilities SKA1-Mid and PUMA. Subsequently, in Sec. \ref{sec:snr}, the projected signal-to-noise ratios (SNR) for the three observables, \emph{i.e.}, Ly$\alpha$ auto, 21 cm auto, and Ly$\alpha$-21 cm cross-power spectra, have been estimated for these instruments corresponding to a range of currently viable fiducial values of $B_0$ and $n_{\rm B}$, which aids us in assessing the detectability of these signals through these upcoming observational facilities. Thereafter, in Sec. \ref{sec:fisher}, Fisher matrix analysis has been performed to robustly forecast on the constraining potential of these different observables when probed via the upcoming experiments with regard to the PMF sector. Finally, we have summarized the important results obtained in our study and concluded by highlighting a few possible future directions to explore in Sec. \ref{sec:concl}.

Throughout the present study, the standard matter power spectrum has been computed using \texttt{CLASS}\footnote{\href{https://github.com/lesgourg/class_public}{https://github.com/lesgourg/class public}}, to which magnetic corrections have subsequently been applied. The Ly$\alpha$ auto-power spectrum and its corresponding noise spectrum have been calculated using \texttt{lyaforecast}\footnote{\href{https://github.com/igmhub/lyaforecast}{https://github.com/igmhub/lyaforecast}}, available via the \texttt{Lyman-$\alpha$ Desi-hub}\footnote{\href{https://github.com/igmhub}{https://github.com/igmhub}}.

\section{PMF-induced matter power at small scales}
\label{sec:pmf_matt}

\begin{figure*}[htbp]
    \centering   
    \subfigure[]{\includegraphics[width=0.48\columnwidth]{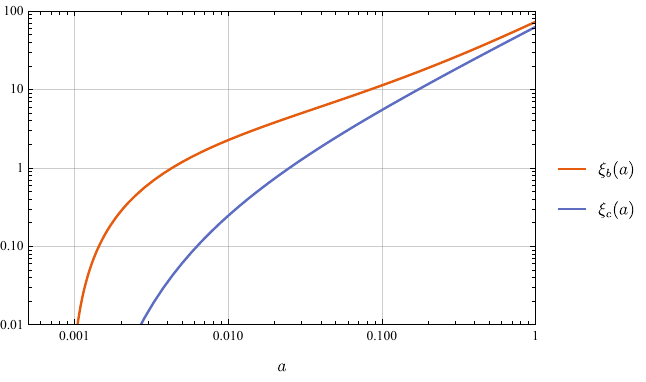}}
    \subfigure[]{\includegraphics[width=0.48\columnwidth]{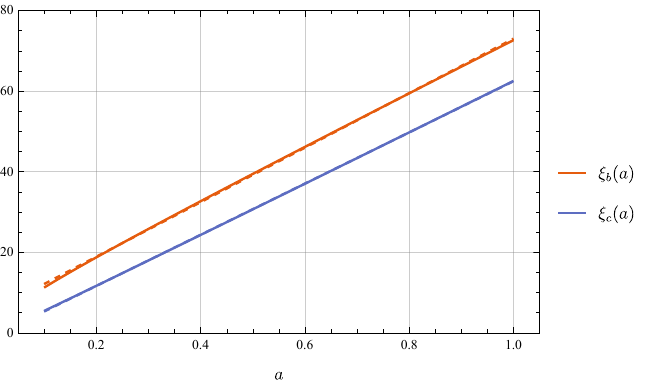}}
    \caption{Growth factors for the PMF-induced baryon and DM perturbations during the post-recombination era as functions of the scale factor, obtained by numerically solving \eqref{eq:baryoneq} and \eqref{eq:dmeq} with zero initial conditions set at recombination. The dashed lines in the right panel show the nearly linear fits at late times, given by the approximate fitting formulae $\xi_b(a)\approx 67.66 a+5.34$ and $\xi_c(a)\approx63.52a-1.08$. The curves are extrapolated to the present time for the purpose of illustration, although the solutions are strictly valid only up to matter-dark energy equality around $a\sim0.75$.}
    \label{fig:magnetic_gr}
\end{figure*}

In this work, we outline the formalism developed in \cite{Ralegankar:2024ekl} that has been adopted in this work to compute the PMF-induced baryon and DM power spectra at small scales. We assume the DM sector to be the standard cold dark matter (CDM) which interacts only gravitationally with the visible matter sector. The Lorentz force term $S(\boldsymbol{x})\propto\nabla .\left[\boldsymbol{B}(\boldsymbol{x})\times\left(\nabla\times\boldsymbol{B}(\boldsymbol{x})\right)\right]$ then sources baryon perturbations, which, in turn, may gravitationally induce the growth of DM perturbations. The starting point is assuming a comoving PMF power spectrum of the power-law type with an exponential damping cut-off, given by
\begin{equation} \label{eq:pmfspectrum}
    P_{\rm B}(k)=A_{\rm B}k^{n_{\rm B}}e^{-k^2\lambda_{\rm D}^2}\:,
\end{equation}
where $A_{\rm B}$ is related to the comoving PMF strength $B_{\rm s}$ smoothed over some given scale $\lambda_s$ as
\begin{equation}
    A_{\rm B}=\dfrac{4\pi^2 B_{\rm s}^2}{\Gamma\left(\frac{n_{\rm B}+3}{2}\right)\lambda_s^{n_{\rm B}+3}}\:.
\end{equation}
For our purpose, we choose $\lambda_s\sim1$ Mpc corresponding to the typical size of the intergalactic voids, which allows us to identify $B_s$ with the PMF magnitude $B_0$ that was introduced earlier. On the other hand, the cut-off scale $\lambda_{\rm D}$ governs the suppression of magnetic power at small scales $k\gtrsim\lambda_{\rm D}^{-1}$ due to turbulent backreaction from baryons \cite{Banerjee:2004df,Trivedi:2018ejz,Ralegankar:2023pyx}. Physically, $\lambda_{\rm D}$ corresponds roughly to the comoving distance covered across Hubble time by a particle moving with the characteristic Alfv\'{e}n speed of the plasma, and it scales identically as the comoving magnetic Jeans scale beyond which magnetic pressure is expected to prevent further gravitational collapse of matter \cite{Kim:1994zh}. A rigorous modeling of $\lambda_{\rm D}$ for generic values of $B_0$ and $n_{\rm B}$ requires magnetohydrodynamical (MHD) simulation, which is beyond the scope of the present work.  However, for nearly scale-invariant PMF spectra having $n_{\rm B}$ close to $-3$, the turbulent damping scale may be well-approximated as \cite{Ralegankar:2024ekl}\footnote{In \cite{Ralegankar:2024ekl}, a proportionality constant, $\kappa_{n_{\rm B}}$, has been introduced on the right hand side of \eqref{eq:lambdeq} to encapsulate the dependence on $n_{\rm B}$. However, semi-analytic estimates of $\kappa_{n_{\rm B}}$ are ambiguous, as they are demonstrably dependent on the choice of an arbitrary $\mathcal{O}(10^{-1})$ number acting as the cut-off on the baryon power spectrum at nonlinear scales, as pointed out in \cite{Ralegankar:2024ekl}. For example, this may lead to differences such as $\kappa_{-2.9}=0.67$ versus $\kappa_{-2.9}=0.88$, or $\kappa_{-2.0}=1.36$ versus $\kappa_{-2.0}=1.79$ \cite{Ralegankar:2024ekl,Ralegankar:2024arh}. In the present study, we thus approximate this $\mathcal{O}(1)$ correction factor in $\lambda_{\rm D}$ with unity, which is not expected to vary significantly anyhow in our case due to our emphasis on closely situated fiducial values of $n_{\rm B}$ in the final forecast analysis.}
\begin{equation} \label{eq:lambdeq}
    \lambda_{\rm D}\approx 0.1\left(\dfrac{B_0}{1\:\textrm{nG}}\right)\:\textrm{Mpc}\:.
\end{equation}
The next necessary step is the computation of the growth factors for the magnetically sourced baryon and DM perturbations, which may be obtained by numerically solving the following set of coupled baryon-DM evolution equations sourced by the Lorentz force:
\begin{equation} \label{eq:baryoneq}
    a^2\dfrac{\partial^2\delta_b}{\partial a^2}+\dfrac{3}{2}a\dfrac{\partial\delta_b}{\partial a}-\dfrac{3}{2}\dfrac{\Omega_{b,0}}{\Omega_{m,0}(1+a_{\rm eq}/a)}\delta_b=-\dfrac{S_0}{a^3H^2}+\dfrac{3}{2}\dfrac{\Omega_{c,0}}{\Omega_{m,0}(1+a_{\rm eq}/a)}\delta_c\:,
\end{equation}
\begin{equation} \label{eq:dmeq}
    a^2\dfrac{\partial^2\delta_c}{\partial a^2}+\dfrac{3}{2}a\dfrac{\partial\delta_c}{\partial a}-\dfrac{3}{2}\dfrac{\Omega_{c,0}}{\Omega_{m,0}(1+a_{\rm eq}/a)}\delta_c=\dfrac{3}{2}\dfrac{\Omega_{b,0}}{\Omega_{m,0}(1+a_{\rm eq}/a)}\delta_b\:.
\end{equation}
Here, $\Omega_{b,0}$ ($\Omega_{c,0}$) denotes the density parameter of baryons (DM) at present time and $\Omega_{m,0}=\Omega_{b,0}+\Omega_{c,0}$, and $a_{\rm eq}\sim2.94\times10^{-4}$ is the scale factor at matter-radiation equality (by normalizing $a_0=1$ today). The dimensionless Lorentz source term is defined as
\begin{equation}
    S_0=-\dfrac{\nabla .\left[\boldsymbol{B}\times\left(\nabla\times\boldsymbol{B}\right)\right]}{4\pi a^3\rho_b(a)}\:,
\end{equation}
where $\rho_b(a)$ is the background baryon density of the Universe. Since $\rho_b(a)\propto a^{-3}$ and the magnetic field $\boldsymbol{B}(\boldsymbol{x})$ is defined to be comoving, $S_0$ is time-independent. Furthermore, during the matter-dominated epoch, $a^3H^2$ is constant. This makes the effective magnetic source term on the right hand side of \eqref{eq:baryoneq} an entirely time-independent quantity. The scale-independent growth factor for each PMF-sourced perturbation is then defined by normalizing as
\begin{equation}
    \xi_b(a)=-\delta_b(a)\left(\dfrac{S_0}{a^3H^2}\right)^{-1},\quad\xi_c(a)=-\delta_c(a)\left(\dfrac{S_0}{a^3H^2}\right)^{-1}\:.
\end{equation}
The behavior of the magnetic growth functions is shown in Fig. \ref{fig:magnetic_gr} by numerically solving the system \eqref{eq:baryoneq} and \eqref{eq:dmeq} from the initial redshift $a_{\rm rec}\sim9.08\times10^{-4}$ up to the present time, with the initial conditions set by assuming $\delta_{b,c}$ and their first derivatives to vanish at $a=a_{\rm rec}$. This framework rests on the underlying assumption that baryon perturbations can be sourced significantly by the PMF only in the post-recombination era, when baryons are no longer tightly coupled to freely streaming photons. At lower redshifts, both $\xi_b(a)$ and $\xi_c(a)$ are seen to grow quasi-linearly and approach each other, which is an expected feature as the DM perturbation traces the baryon perturbation more efficiently at later times. 

With the PMF power spectrum and the magnetic growth factor at hand, the PMF-induced baryon power spectrum at small scales is given by 
\begin{equation} \label{eq:pbaryon}
    P_{\rm b,PMF}(k,a)=\dfrac{k^4\xi_b(a)^2}{128\pi^2a^6(a^3H^2)^2\rho_b(a)^2}\int \dfrac{d^3q}{(2\pi)^3}P_{\rm B}(q)P_{\rm B}(|\boldsymbol{k}-\boldsymbol{q}|)\mathcal{G}(v,\theta)\:,
\end{equation}
with the angular part of the integrand defined as
\begin{equation}
    \mathcal{G}(v,\theta)=\dfrac{1+\cos^2\theta+2v^2(1-2\cos^2\theta+2\cos^4\theta)-4v\cos^3\theta}{1+v^2-2v\cos\theta}\:,
\end{equation}
where $v=q/k$ and $\cos\theta=\frac{\boldsymbol{k}.\boldsymbol{q}}{kq}$. Using the form of the PMF spectrum from \eqref{eq:pmfspectrum}, for $-3<n_{\rm B}<-1.5$, the dimensionless baryon power spectrum can then be well-approximated as
\begin{equation}
    \Delta_{\rm b,PMF}^2(k,a)=\dfrac{k^3}{2\pi^2}P_{\rm b,PMF}(k,a)\approx 10^{-4}\xi_b(a)^2\left(\dfrac{k}{1\:\textrm{Mpc}^{-1}}\right)^{2n_{\rm B}+10}\left(\dfrac{B_0}{1\:\textrm{nG}}\right)^4\mathcal{F}(k,\lambda_{\rm D},n_{\rm B})\:,
\end{equation}
where the numerical pre-factor originates from substituting the values of $\rho_{b,0}\approx1.64\times10^{-48}\:\textrm{GeV}^4$ and $\rho_{m,0}\approx1.15\times10^{-47}\:\textrm{GeV}^4$ into \eqref{eq:pbaryon}, and the scale-dependent angular contribution is expressed as
\begin{eqnarray}
    \mathcal{F}(k,\lambda_D,n_{\rm B})=&& \dfrac{1}{2\Gamma\left(\frac{n_{\rm B}+3}{2}\right)^2}\int\limits_0^\infty dy\int\limits_{-1}^{1} y^{n_{\rm B}+2}(1+y^2-2yx)^{\frac{n_{\rm B}-2}{2}} \nonumber \\ 
    &&\times \left(1+2y^2+4y^2x^4-4y^2x^2-4yx^3+x^2\right)\exp\left[-\lambda_{\rm D}^2k^2(1+2y^2-2yx)\right]\:.
\end{eqnarray}
Following \cite{Pavicevic:2025gqi}, the total matter power spectrum, $\Delta_{\rm m,PMF}^2$, can then be related to the baryon power spectrum by means of growth functions as
\begin{equation}
    \Delta_{\rm m,PMF}^2(k,a)=\left[\dfrac{\Omega_{b,0}}{\Omega_{m,0}}+\dfrac{\xi_c(a)}{\xi_b(a)}\dfrac{\Omega_{c,0}}{\Omega_{m,0}}\right]^2\Delta_{\rm b,PMF}^2(k,a)\:.
\end{equation}
The total matter power spectrum may then be estimated as a linear sum of the standard matter power in the usual $\Lambda$CDM scenario and the PMF-induced contribution as
\begin{equation}
    \Delta_{\rm m,tot}^2(k,a)=\Delta_{\rm m,\Lambda CDM}^2(k,a)+\Delta_{\rm m,PMF}^2(k,a)\:,
\end{equation}
where cross-correlations between standard matter fluctuations and PMF-induced matter perturbations have been neglected. For our purpose, the standard matter power spectrum has been computed using the Einstein-Boltzmann solver \texttt{CLASS} \cite{2011arXiv1104.2932L,2011JCAP...07..034B} corresponding to the best-fit values for the $\Lambda$CDM parameters from \textit{Planck} 2018 data  \cite{Planck:2018vyg}. 

\begin{figure*}[htbp]
    \centering   
    \subfigure[]{\includegraphics[width=0.48\columnwidth]{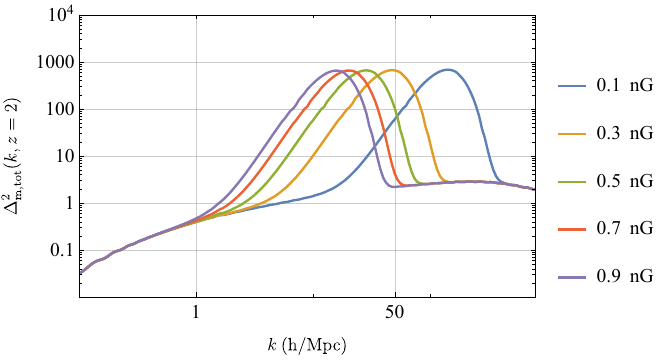}}
    \subfigure[]{\includegraphics[width=0.48\columnwidth]{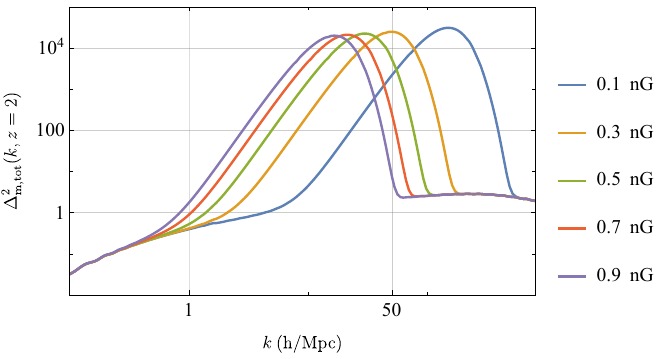}}
    \quad
    \subfigure[]{\includegraphics[width=0.48\columnwidth]{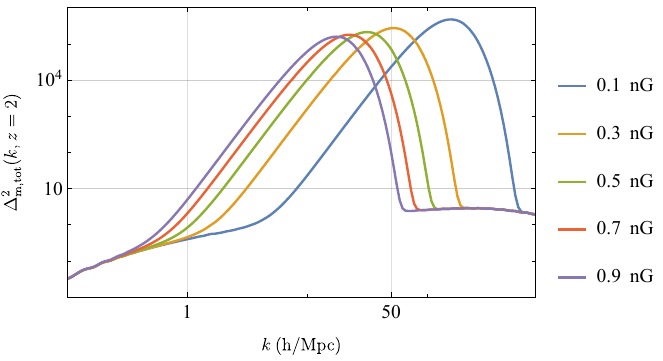}}
    \subfigure[]{\includegraphics[width=0.48\columnwidth]{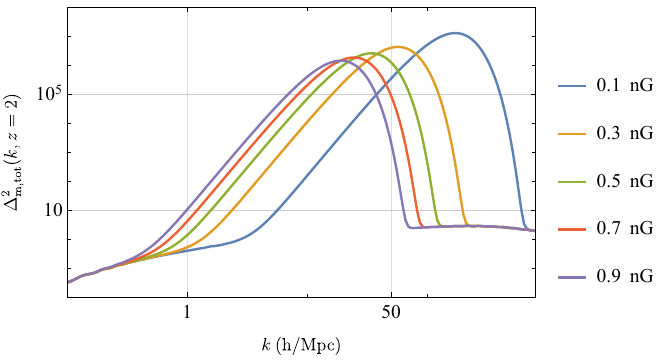}}
    \caption{The total dimensionless matter power spectrum, given by the sum of the standard $\Lambda$CDM matter power spectrum and the PMF-induced matter power spectrum at small scales, computed at $z=2.0$ for a few representative sub-nG values of $B_0$ and four values of the magnetic spectral index: $n_B=-2.99$ (a), -2.90 (b), -2.75 (c), and -2.50 (d).}
    \label{fig:magnetic_panels}
\end{figure*}

\section{Modeling the Ly$\alpha$ and 21-cm observables} \label{sec:pmf_ly_21}

In this section, we outline the theoretical modeling of the Ly$\alpha$ and 21 cm power spectra and the instrumental noise models which are used in our subsequent forecast analysis. Our adopted semi-analytic approach relies on a fiducial matter power spectrum enhanced locally at small scales by a PMF that is provided as the input. The Ly$\alpha$ and 21 cm fluctuations may then be modeled as biased tracers of the total matter perturbation, with the bias functions being both redshift and scale-dependent at the nonlinear scales of interest $k\gtrsim1.0\:h/\textrm{Mpc}$, where the enhancement of matter power typically occurs for the PMF parameters in the ballpark ranges of $B_0\lesssim 1$ nG and $-3<n_{\rm B}<-1.5$.

\subsection{Ly$\alpha$ power spectrum} \label{subsec:ly}

The 3D Ly$\alpha$ flux power is a biased tracer of the underlying total matter power spectrum, and can be expressed as
\begin{equation}
    P_F^{\rm 3D}(\boldsymbol{k},z)=b_F^2\left(1+\beta_F\mu^2\right)^2P_{\rm m,tot}(k,z)\:,
\end{equation}
where $\mu\equiv k_{\parallel}/k$, and we use the approximate fitting forms of the flux bias, $b_F$, and the linear redshift space distortion (RSD) parameter, $\beta_F$, from \cite{Arinyo-i-Prats:2015vqa} along with the redshift extension factor provided in \cite{BOSS:2013rpr}. This term represents the contribution from cosmic variance, \emph{i.e.}, it is given by the two-point correlator of the Ly$\alpha$ flux fluctuations $\langle\delta_F\delta_F^*\rangle$ in Fourier space.

The total power spectrum, $P_F^{\rm T}(\boldsymbol{k},z)$, may then be decomposed as the sum of three distinct terms as follows \cite{McDonald:2006qs}:
\begin{equation}
    P_F^{\rm T}(\boldsymbol{k},z)=P_F^{\rm 3D}(\boldsymbol{k},z)+P_F^{\rm 1D}(k_{\parallel},z)P_w^{\rm 2D}(z)+P_N^{\rm eff}(z)\:.
\end{equation}
In the second term, the 1D flux power spectrum measured along the line of sight is denoted by $P_F^{\rm 1D}(k_{\parallel},z)$, while $P_w^{\rm 2D}(z)$ encapsulates the contribution of weights assigned to spatial locations in the transverse direction to account for the random distribution of quasars. The second term thus amounts overall to the aliasing noise originating from the sparse sampling of QSOs, and may be computed according to the detailed analytic pipeline given in \cite{McDonald:2006qs}. The third term is the effective noise of the spectrographic instrument, which we discuss in detail for the scenario of our choice in a subsequent section. Following \cite{Villaescusa-Navarro:2014rra,Carucci:2016yzq}, the variance of the total Ly$\alpha$ power spectrum can then be expressed as
\begin{equation} \label{eq:sigF}
    \sigma^2\left[P_F(\boldsymbol{k},z)\right]=\left[P_F^{\rm T}(\boldsymbol{k},z)\right]^2\:,
\end{equation}
which helps assess the SNR for the Ly$\alpha$ power spectrum at an instrument whose noise response function, $P_N^{\rm eff}$, is known.

For the Ly$\alpha$ sector, we consider a next-generation DESI-like spectroscopic survey mission designed to achieve $\frac{dn}{dmdzd\Omega}\sim100$, where $n$ is the number density of detected Ly$\alpha$ QSO samples, $m$ is the apparent magnitude, $z$ is the redshift, and $\Omega$ is the solid angle on the sky plane (expressed in deg\textsuperscript{2}). The expected number density of QSO samples is determined by the quasar luminosity function (QLF) \cite{Hopkins:2005ca,2013A&A...551A..29P,Yeche:2017upn,2023ApJ...949L..42M}.

\subsection{21 cm power spectrum} \label{subsec:21}

In the post-EoR era, the 3D power spectrum of the 21 cm fluctuations, $P_{\rm 21}(\boldsymbol{k},z)$, is a biased tracer of the matter power spectrum, and can be expressed as
\begin{equation}
    P_{21}(\boldsymbol{k},z)=\bar{T}(z)^2b_{\rm HI}^2\left(1+\beta_{\rm HI}\mu^2\right)^2P_{\rm m,tot}(k,z)\:,
\end{equation}
with the global 21 cm brightness temperature, $\bar{T}(z)$, being given by~\cite{CosmicVisions21cm:2018rfq,Obuljen:2017jiy,Obuljen:2018kdy}
\begin{equation}\label{eq:Tbar}
    \bar{T}(z) = 188\,\frac{h}{E(z)}\,\Omega_{\rm HI}(z)\,(1+z)^2\;\rm mK,
\end{equation}
where $E(z) = H(z)/H_0$ and $\Omega_{\rm HI}(z) = 4\times 10^{-4}(1+z)^{0.6}$ is the HI density parameter. 

At mildly non-linear scales, the spatial distribution of HI is governed by a scale-dependent, complex bias. We approximate the non-linear HI bias function analytically as a mixed polynomial following~\cite{Sarkar:2016lvb}:
\begin{equation}
    b_{\mathrm{HI}}(k,z) = \sum_{m=0}^4\sum_{n=0}^2 c(m,n)k^mz^n\:.
\end{equation}
Regarding its domain of validity, the coefficients $c(m,n)$ derived from baseline hydrodynamical simulations~\cite{Sarkar:2016lvb} yield a $b_{\mathrm{HI}}(k,z)$ that is robustly scale-independent at linear scales ($k \lesssim 0.1 \, \mathrm{Mpc}^{-1}$), but exhibits significant complex stochasticity at highly non-linear scales. The underlying simulations assume a minimum halo mass of $10^9 M_{\odot}$, which accurately captures the HI distribution at $z \leq 3.5$ but may slightly underestimate low-mass halo contributions at higher redshifts. The best-fit values of the polynomial coefficients are listed below, as obtained by \cite{Sarkar:2016lvb}:
\begin{equation}
    c(m,n)\times10^{-2}=\begin{pmatrix} $65.31$ & $25.19$ & $1.963$ \\ $-60.74$ & $18.56$ & $1.806$ \\ $33.54$ & $-17.38$ & $1.618$ \\ $-5.129$ & $3.247$ & $-0.3803$ \\ $0.2773$ & $-0.1899$ & $0.02435$ \end{pmatrix}\:,
\end{equation}
where the rows correspond to $0<m<4$ and the columns correspond to $0<n<2$, and the associated $1\sigma$ errors are $\lesssim10\%$. These fitting forms hold for $k\lesssim20\:h/\textrm{Mpc}$ across the redshift range $z\sim2-6$.

The total power spectrum, $P_{21}^T(\boldsymbol{k},z)$, is subsequently given by the sum of the 3D power spectrum (cosmic variance) and the total noise, whose modeling depends on whether the HI IM survey instrument is designed to operate in the interferometric (IF) mode or in the single dish (SD) mode. Schematically, the variance of the 21 cm power spectrum may then be expressed as
\begin{equation} \label{eq:sig21}
    \sigma^2\left[P_{21}(\boldsymbol{k},z)\right]=\left[P_{21}^T(\boldsymbol{k},z)\right]^2=\left[P_{21}(\boldsymbol{k},z)+N_{21}(k, \mu, z)\right]^2\:
\end{equation}
The total noise in the 21\,cm auto-power spectrum receives contributions from both thermal noise in the receiver and Poisson shot noise arising from the discrete nature of the HI source distribution. The total noise power spectrum is~\cite{Bull:2014rha,Castorina:2016bfm,Karagiannis:2020dpq,Jolicoeur:2020eup,Aharonian:2013av,Bonaldi:2025gpa}
\begin{equation}
    N_{21}(k, \mu, z) = N_{\rm thermal}(k, \mu, z) + \bar{T}^2(z)\,P_{\rm shot}(z),
\end{equation}
where $\bar{T}(z)$ is the sky-averaged HI brightness temperature defined in Eq.~\ref{eq:Tbar} and $P_{\rm shot}(z)$ is the Poisson shot noise in $({\rm Mpc}/h)^3$. Both $P_{\rm shot}(z)$ and the HI bias $b_{\rm HI}(z, k)$ are taken from the tabulation ~\cite{Castorina:2016bfm}, interpolated as smooth functions of redshift. The two-dimensional nature of the noise, however, depends on whether the instrument operates in interferometer mode or single-dish mode, and these two cases are treated separately below.
\subsubsection{PUMA (Interferometer Mode)} \label{subsubsec:pumamodel}
A proposed futuristic successor of the SKA1 facility, PUMA is designed to operate in both SD mode \cite{Karagiannis:2020dpq} and IF mode \cite{Jolicoeur:2020eup}. It is envisioned to be a hexagonally close-packed array of dishes with a bandwidth of $200-1100$ MHz, allowing it to probe the approximate redshift range of $0.3<z<6$ \cite{CosmicVisions21cm:2018rfq}. In interferometer mode, the instrument records the complex visibilities formed by correlating pairs of antenna elements. The thermal noise in the reconstructed power spectrum depends on the baseline density distribution of the array, which determines how many independent measurements contribute to each Fourier mode in the transverse plane. For a baseline $\mathbf{u}$ in the image plane, the effective number density of baselines at that location is $n(u)$, where $u = |\mathbf{u}|$. For a given transverse wavenumber $k_\perp$, the relevant baseline length is $u = k_\perp \chi(z)/(2\pi)$, where $\chi(z)$ is the comoving angular diameter distance. The thermal noise power spectrum in the interferometer mode is \citep{Karagiannis:2020dpq,Jolicoeur:2020eup}
\begin{equation}
    N_{\rm thermal}^{\rm IF}(k_\perp, z) = T_{\rm sys}^2(z)\,\frac{\chi^2(z)\,y(z)}{t_{\rm int}} \cdot \frac{\lambda^4(z)}{A_e^2} \cdot \frac{1}{2\,n(u, z)} \cdot \frac{S_{\rm area}}{\theta_{\rm FOV}(z)^2},
\end{equation}
where $\lambda(z) = 0.21(1+z)\,\rm m$ is the observed wavelength of the 21\,cm line, $A_e = (\pi/4)D^2\eta$ is the effective collecting area of a single dish of physical diameter $D$ and aperture efficiency $\eta$, $t_{\rm int}$ is the total on-sky integration time\footnote{Note that $t_{\rm int}$ is essentially distinct from the observation time of the instrument, $t_{\rm obs}$, in the sense that the former represents the summed contribution of the timescales over which data is collected and integrated by all the antennae. As such, $t_{\rm int}$ is estimated by accounting for instrumental parameters such as the baseline and effective collecting area of each antenna as well as the total number of operating dishes (\emph{e.g.}, see \cite{Aharonian:2013av,SKA:2018ckk,Bonaldi:2025gpa}).}, $S_{\rm area}$ is the survey solid angle, and $\theta_{\rm FOV}(z) = 1.22 \,\lambda(z)/D_{\rm Dish}$ is the primary beam field of view. The factor $y(z) = c(1+z)^2/(\nu_{21}H(z))$ converts a frequency bandwidth into a comoving line-of-sight depth. The factor of 2 in the denominator accounts for the two polarization channels measured by each antenna element.

The baseline density $n(u, z)$ encodes the array configuration and determines the noise as a function of transverse scale. For a hexagonally close-packed array with $N_{\rm side}$ dishes per side, each of diameter $D$, the baseline density is well described by the empirical fitting function~\cite{CosmicVisions21cm:2018rfq,PUMA:2019jwd}
\begin{equation}
    n(u) = n_0\,\frac{a + b\,\tilde{u}}{1 + B\,\tilde{u}^{\,C}} \exp\!\left(-\tilde{u}^{\,D}\right), \qquad \tilde{u} = \frac{u}{N_{\rm side}\,D}, \qquad n_0 = \left(\frac{N_{\rm side}}{D}\right)^2,
\end{equation}
where the coefficients $(a, b, B, C, D)$ take different values for square-packed and hexagonally close-packed layouts. For the hexagonal packing appropriate to PUMA, the fitted values are $a = 0.5698$, $b = -0.5274$, $B = 0.8358$, $C = 1.6635$, $D = 7.3178$.

The system temperature entering the noise expression is
\begin{equation}
    T_{\rm sys}(z) = T_{\rm sky}(\nu) + T_{\rm scope},
\end{equation}
where the sky temperature is dominated by Galactic synchrotron emission and follows the power-law relation $T_{\rm sky}(\nu) = 25(\nu/400\,{\rm MHz})^{-2.75} + 2.75\,\rm K$, and the receiver contribution is $T_{\rm scope} = T_{\rm ampl}/\eta_{\rm feed}^2 + T_{\rm ground}(1 - \eta_{\rm feed})/\eta_{\rm feed}$, encoding both the amplifier noise temperature $T_{\rm ampl}$ and the ground spillover characterized by the feed efficiency $\eta_{\rm feed}$.

In interferometer mode, only a finite range of transverse wavenumbers is accessible, set by the minimum and maximum baselines of the array. Modes outside the range $k_{\min} = 0.01\,h\,{\rm Mpc}^{-1}$ and $k_{\max} = 5\,h\,{\rm Mpc}^{-1}$ are excluded from the Fisher analysis.
The instrumental specifications adopted for PUMA in this work are summarized in Table~\ref{tab:puma_specs}.

\begin{table}[h]
    \centering
    \renewcommand{\arraystretch}{1.3}
    \begin{tabular}{lc}
        \hline\hline
        Parameter & Value \\
        \hline
        Operating mode & Interferometer (hexagonal close-packed) \\
        Number of dishes per side, $N_{\rm side}$ & 256 \\
        Dish diameter, $D$ & 6\,m \\
        Aperture efficiency, $\eta$ & 0.7 \\
        Integration time, $t_{\rm int}$ & 5\,yr \\
        Sky fraction, $f_{\rm sky}$ & 0.5 \\
        Amplifier temperature, $T_{\rm ampl}$ & 50\,K \\
        Ground temperature, $T_{\rm ground}$ & 300\,K \\
        \hline\hline
    \end{tabular}
    \caption{Instrumental specifications for the PUMA interferometer configuration adopted in this work.}
    \label{tab:puma_specs}
\end{table}
\subsubsection{SKA1-Mid (Single-Dish Mode)} \label{subsubsec:ska1model}
The SKA1-Mid survey is designed to operate in the SD mode across two frequency bands, with SKA1-Mid Band 1 covering the redshift range $0.35<z<3.05$ and SKA1-Mid Band 2 covering $0.10<z<0.49$ \cite{SKA:2018ckk,Jolicoeur:2020eup}. In single-dish mode, each dish operates independently as a total-power radiometer, mapping the sky by scanning across it. The individual maps from all dishes are subsequently co-added, reducing the thermal noise by a factor $\sqrt{N_{\rm dish}}$ relative to a single dish. Unlike the interferometer case, the thermal noise in single-dish mode is approximately isotropic in Fourier space to leading order, since there is no baseline density weighting by transverse scale. The thermal noise power spectrum is~\cite{Bull:2014rha,Karagiannis:2020dpq}
\begin{equation}
    N_{\rm thermal}^{\rm SD}(z) = \frac{T_{\rm sys}^2(z)\,\chi^2(z)\,y(z)\,S_{\rm area}}{2\,N_{\rm dish}\,t_{\rm survey}},
\end{equation}
where $N_{\rm dish}$ is the total number of dishes, $t_{\rm survey}$ is the survey integration time, and the factor of 2 again accounts for dual-polarization measurements. The system temperature retains the same form as in the interferometer case, with $T_{\rm sys} = T_{\rm sky}(\nu) + T_{\rm rx}$, where $T_{\rm rx}$ is the receiver noise temperature of each dish.

The angular resolution of a single dish sets an upper limit on the accessible transverse wavenumbers. Modes with $k_\perp$ beyond the beam scale are exponentially suppressed in sensitivity. Rather than multiplying the noise expression by an explicit beam window function, we follow the approach of imposing sharp cuts on the accessible transverse scale range~\citep{Bull:2014rha}, where we set $k_{\rm min}=0.01 \, h\,{\rm Mpc}^{-1}$ and $k_{\rm max}= 10 \, h\, {\rm Mpc}^{-1}$.
All modes with $k_\perp$ outside this range are excluded from the Fisher analysis. The SKA-Mid Band~1 specifications adopted in this work are given in Table~\ref{tab:ska_specs}.

\begin{table}[h]
    \centering
    \renewcommand{\arraystretch}{1.3}
    \begin{tabular}{lc}
        \hline\hline
        Parameter & Value \\
        \hline
        Operating mode & Single dish \\
        Number of dishes, $N_{\rm dish}$ & 197 \\
        Dish diameter, $D_{\rm dish}$ & 15\,m \\
        Survey area, $S_{\rm area}$ & 20\,000\,deg$^2$ \\
        Total integration time, $t_{\rm int}$ & 10\,000\,hr \\
        Receiver temperature, $T_{\rm rx}$ & 20\,K \\
        Aperture efficiency, $\eta$ & 1.0 \\
        \hline\hline
    \end{tabular}
    \caption{Instrumental specifications for the SKA-Mid Band~1 single-dish configuration adopted in this work.}
    \label{tab:ska_specs}
\end{table}

\subsection{Ly$\alpha$-21 cm cross-power spectrum} \label{subsec:lya21cross}

The third important observable that we focus on  results from the cross-correlation between the Ly$\alpha$ and 21 cm fluctuations, usually quantified by  
\begin{eqnarray}
    P_{21,F}(\boldsymbol{k},z)=b_Fb_{\rm HI}\left(1+\beta_F\mu^2\right)\left(1+\beta_{\rm HI}\mu^2\right)P_{\rm m,tot}(k,z)\:,
\end{eqnarray}
which is  the cross-power spectrum of the Ly$\alpha$ flux and 21 cm brightness temperature fluctuations. As mentioned in Sec. \ref{sec:intro}, it is important to note that this cross-correlation is a biased tracer of the matter power spectrum despite its two constituent fluctuations originating from very distinct physical processes.  The variance of the cross-correlation signal may subsequently be expressed as
\begin{equation}
    \sigma^2\left[P_{21,F}(\boldsymbol{k},z)\right]=\dfrac{1}{2}\left[P_{21,F}(\boldsymbol{k},z)^2+\sigma\left[P_{21}(\boldsymbol{k},z)\right]\sigma\left[P_F(\boldsymbol{k},z)\right]\right]\:,
\end{equation}
with the respective variances of the autocorrelation signals being given by \eqref{eq:sigF} and \eqref{eq:sig21}.

A critical aspect of modeling the variance for the Ly$\alpha$-21 cm cross-correlation is the consideration of a common sky coverage area for the spectroscopic and the radio facilities, whose synergy would, in principle, enable the detection of this signal. In a realistic scenario, the separated geographical locations of the two missions are expected to play a key role in determining the sky overlap, \emph{e.g.}, the fact that spectroscopic large scale survey instruments like DESI are traditionally based in the northern hemisphere (barring a few specific proposals like the 4-metre Multi-Object Spectroscopic Telescope \cite{Richard:2019dwt}), whereas radio facilities like the SKA would operate primarily in the southern hemisphere, should result in a reduced sky overlap compared to their individual coverage areas. This reduction must be taken into account for a proper assessment of the detectability of the Ly$\alpha$-21 cm cross-signal based on the synergy of such instruments. We quantitatively address this issue in Sec. \ref{sec:snr} while estimating the SNR of the cross-correlation signal for suitable combinations of next-generation spectroscopic and radio-interferometric survey missions, which are interestingly seen to allow sufficiently high levels of SNR for the cross-correlation that are competitive with the auto-spectra in spite of its relatively smaller net sky coverage area.

One more crucial reason for considering the Ly$\alpha$-21 cm cross-spectrum, as demonstrated in \cite{Carucci:2016yzq} is that, it is expected to be much less contaminated by foregrounds compared to the 21 cm auto-spectrum, which makes the former a valuable post-EoR probe of matter clustering at small cosmological scales $k\gtrsim1\:h/\textrm{Mpc}$. Our present study is optimistic in the sense that we neglect the presence of foregrounds in our analysis, which implicitly relies on the assumption that the cosmological 21 cm power spectrum may be efficiently disentangled from astrophysical and terrestrial foregrounds. Under such a premise, if the Ly$\alpha$-21 cm cross-spectrum and the 21 cm auto-spectrum exhibit comparable levels of projected SNR and error forecasts on model parameters at future instruments, it may be prudent to surmise that the Ly$\alpha$-21 cm cross-spectrum could overall be a better observable in terms of both detectability and constraining power once realistic foregrounds are taken into account.

\section{Estimation of signal-to-noise ratio (SNR)} \label{sec:snr}

Having set the stage by decomposing each observable signal as the sum of its fiducial theoretical component and the instrumental noise power, we now proceed towards the estimation of the projected SNR for the Ly$\alpha$ auto, 21 cm auto, and Ly$\alpha$-21 cm cross-power spectra, to be probed via the next-generation spectroscopic and radio survey missions described in Sec. \ref{sec:pmf_ly_21}. 

Given a signal, $P_i(\boldsymbol{k},z)$, and its variance, $\sigma\left[P_i(\boldsymbol{k},z)\right]$, the SNR can be expressed as
\begin{equation} \label{eq:snr}
    \textrm{SNR}_i^2=\dfrac{V_{\rm survey}(z_c)k^3\epsilon d\mu}{4\pi^2}\dfrac{P_i(\boldsymbol{k},z)^2}{\sigma^2\left[P_i(\boldsymbol{k},z)\right]}\:,
\end{equation}
where $\epsilon\equiv dk/k$ is the logarithmic width of the $k$-bin, and $V_{\rm survey}(z_c)$ is the comoving annular survey volume for a redshift bin centered at redshift $z_c$ defined as
\begin{equation}
    V_{\rm survey}(z_c)=\dfrac{4\pi}{3}f_{\rm sky}\left[\chi(z_{\rm max})^3-\chi(z_{\rm min})^3\right]\:,
\end{equation}
where $z_{\rm min}$ and $z_{\rm max}$ are the two redshift boundaries for the $z_c$-centered bin, determined by the frequency bandwidth of the instrument, $\Delta\nu$. Equipped with the results from Sec. \ref{sec:pmf_ly_21}, the expression in \eqref{eq:snr} then allows computing the SNR values for all three observables ($i=$ Ly$\alpha$ auto, $21$ cm auto, and Ly$\alpha$-21 cm cross) of our interest. Importantly, the SNR is generically a function of the wavenumber ($k$), the redshift ($z$), and the direction cosine ($\mu$). We present the SNR by keeping the dependence on all three of these arguments explicit, which aids in a transparent comparison among its general trends.

Although the sky survey areas for both the DESI-like spectroscopic survey and the 21 cm array (SKA1-Mid/PUMA) would individually be of order $\sim 10^4$ deg\textsuperscript{2}, their net sky overlap is what determines the prospect of detecting the Ly$\alpha$-21 cm cross-correlation signal, as discussed earlier in Sec. \ref{subsec:lya21cross}.  We choose $S_{\rm area}=1000$ deg\textsuperscript{2} for the cross-spectrum similar to \cite{Montero-Camacho:2024xvf}, whereas for the Ly$\alpha$ and 21 cm auto spectra, we respectively take $S_{\rm area}=1.5\times10^4$ deg\textsuperscript{2} and $S_{\rm area}=2.0\times10^4$ deg\textsuperscript{2}. We also fix the integration time  for the 21-cm surveys at a conservative value of $t_{\rm int}=400$ hours, which is significantly smaller than their projected integration timescales of order $\sim10^4$ hours. Our purpose in this study is to demonstrate that even such modest instrumental choices for next-generation detectors could lead to considerably tight constraints on the PMF sector in the light of post-EoR observables. 

\begin{figure*}[htbp]
    \centering
    \includegraphics[width=1.0\textwidth]{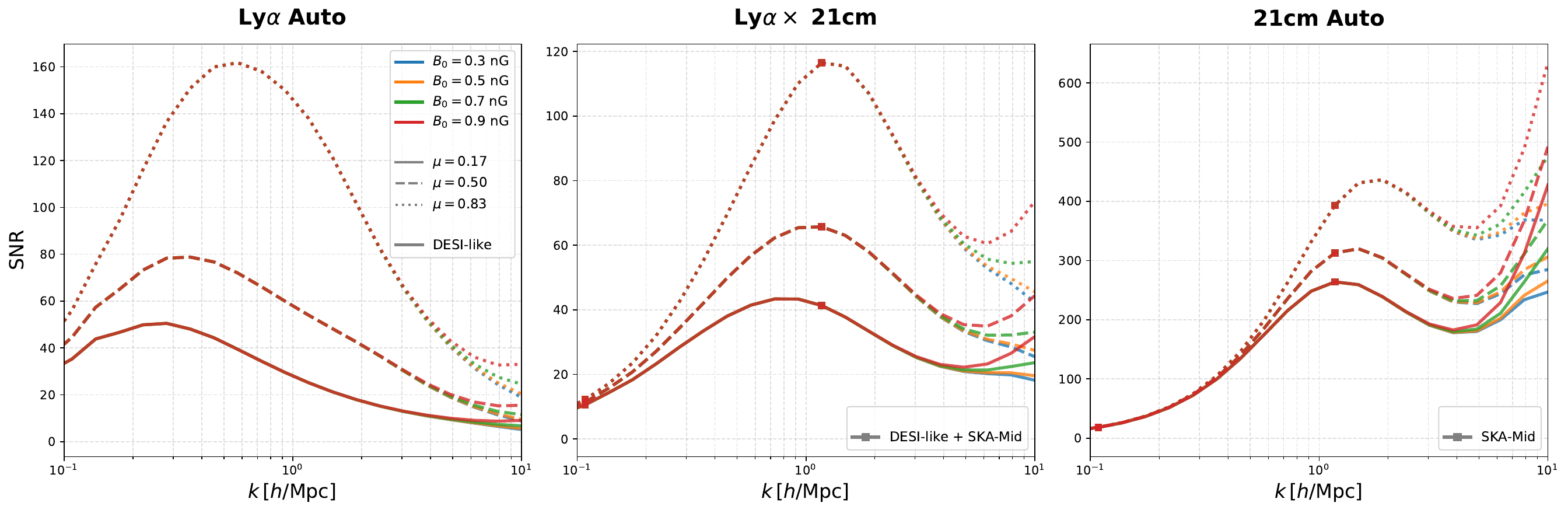}
    \includegraphics[width=1.0\textwidth]{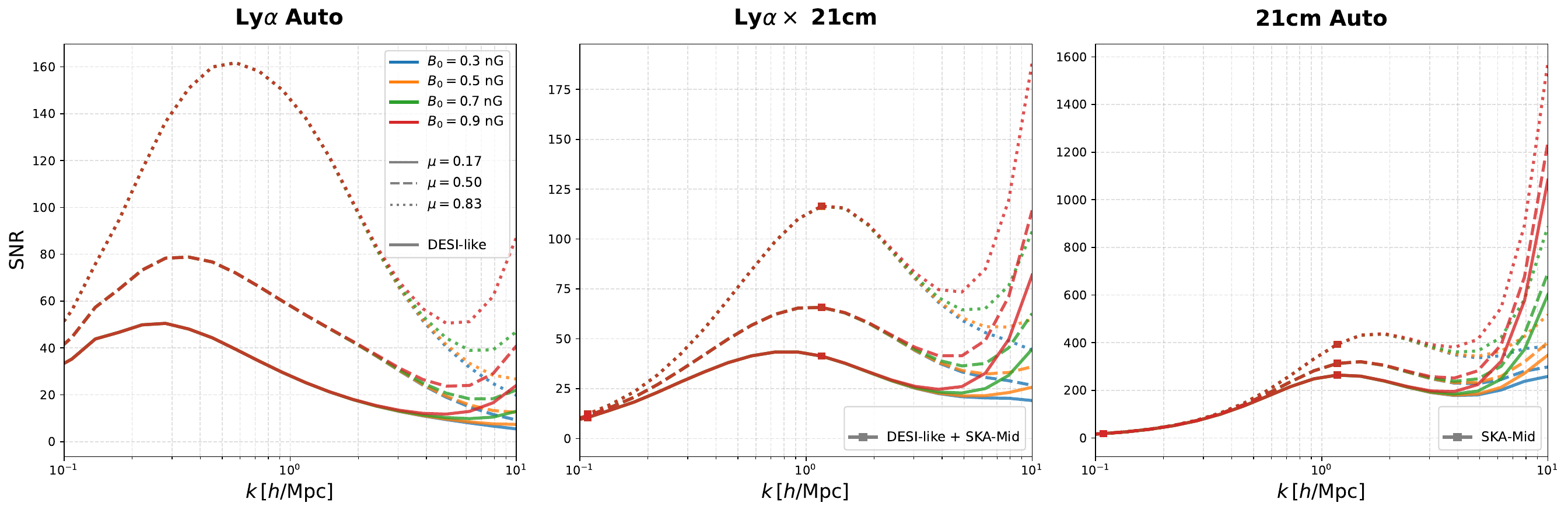}
    \caption{Signal-to-noise ratio (SNR) for the Ly$\alpha$ auto, the 21 cm auto, and the Ly$\alpha$-21 cm cross spectra shown as a function of scale, in presence of a PMF with $n_B=-2.9$ (\emph{upper panel}) and $n_{\rm B}=-2.75$ (\emph{lower panel}) and a few benchmark values of $B_0\lesssim1$ nG as mentioned in the plots. The results are shown at a fixed redshift $z=2.33$ for a few fixed values of $\mu$. For the instruments, we assume the DESI-like spectroscopic detector (for Ly$\alpha$) and the SKA1-Mid configuration (for 21 cm), as described in Sec. \ref{subsubsec:ska1model}.}
    \label{fig:snrvsk_ska}
\end{figure*}

\begin{figure*}[htbp]
    \centering
    \includegraphics[width=1.0\textwidth]{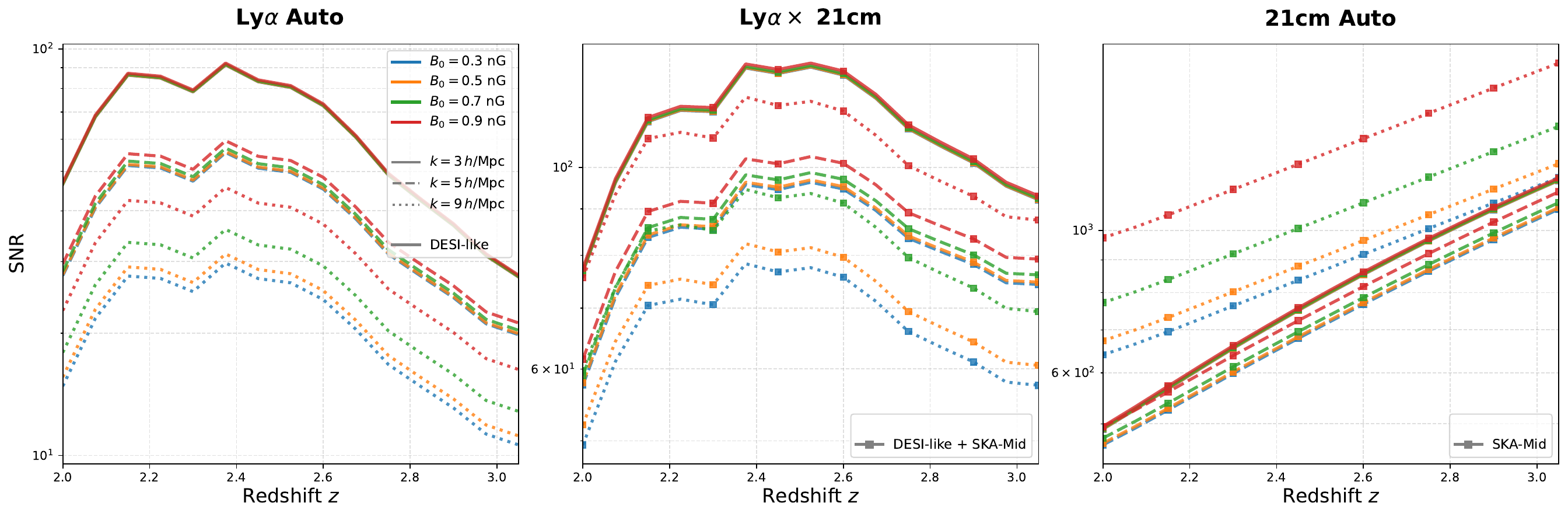}
    \includegraphics[width=1.0\textwidth]{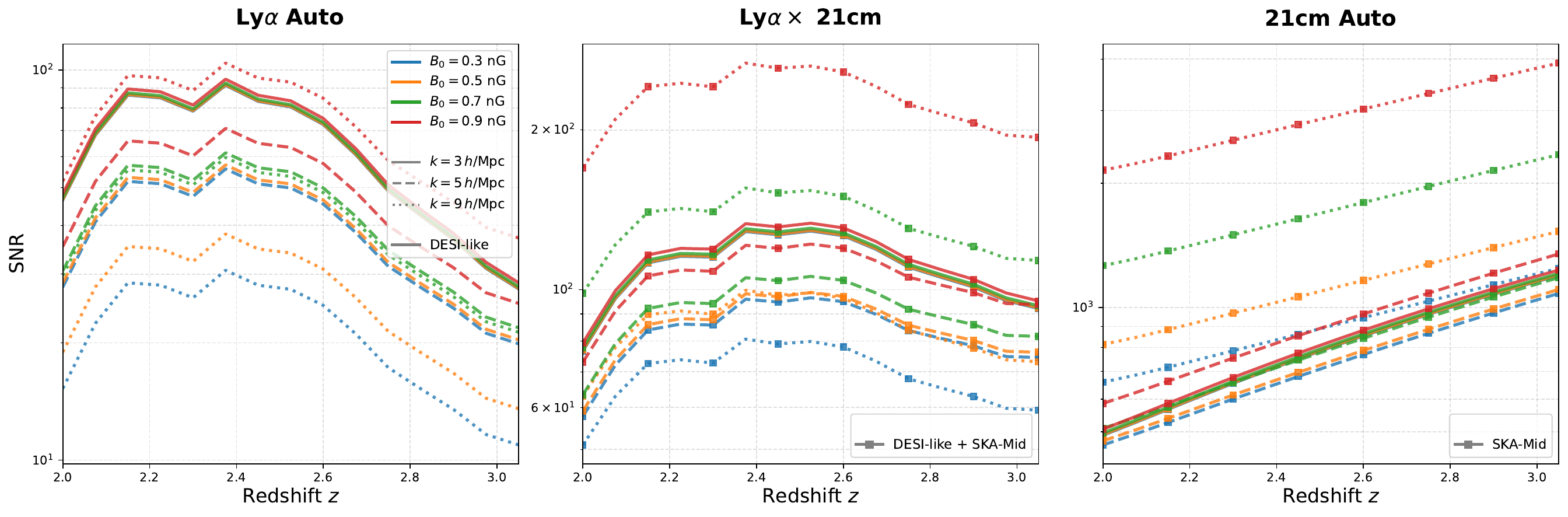}
    \caption{Signal-to-noise ratio (SNR) for the Ly$\alpha$ auto, the 21 cm auto, and the Ly$\alpha$-21 cm cross spectra shown as a function of redshift, in presence of a PMF with $n_B=-2.9$ (\emph{upper panel}) and $n_{\rm B}=-2.75$ (\emph{lower panel}), and a few benchmark values of $B_0\lesssim1$ nG as mentioned in the plots. The different curves correspond to a few benchmark values of the scale as mentioned in the plots, and $\mu=0.5$ has been fixed for the purpose of illustration. For the instruments, we assume the DESI-like spectroscopic detector (for Ly$\alpha$) and the SKA1-Mid configuration (for 21 cm), as described in Sec. \ref{subsubsec:ska1model}.}
    \label{fig:snrvsz_ska}
\end{figure*}

For the DESI-like spectroscopic survey instrument and the SKA1-Mid radio facility (SD mode) described in Sec. \ref{subsubsec:ska1model}, the SNR for the three observables is shown as a function of scale and of redshift respectively in Figs. \ref{fig:snrvsk_ska} and \ref{fig:snrvsz_ska}, for a set of comoving PMF amplitudes of sub-nG strength and two representative values $n_{\rm B}=-2.9$ and $n_{\rm B}=-2.75$ of the magnetic spectral index. For the DESI-like and SKA1-Mid combination, the shortest scale within reach has been fixed at  $k=10h/\textrm{Mpc}$, and the redshift range is chosen to be $2.00<z<3.05$ which falls within the specifications for SKA1-Mid Band 1 (see Table \ref{tab:ska_specs}). For the three different signals that are considered, we then observe the following general trend for the overall SNR values\footnote{It is important to recall that our present analysis neglects the impact of foregrounds, and as such, is based on a fairly simplistic noise-modeling scheme that takes into account only cosmic variance, thermal noise, and instrumental noise. This, alongside a comparatively larger sky coverage area, allows the 21 cm auto-spectrum to apparently dominate over the Ly$\alpha$-21 cm cross-spectrum in terms of the SNR. However, in the presence of realistic foregrounds, the inequalities may not strictly hold, which is subject to arrival of real data in future.
 }: 
\begin{equation}
    \textrm{SNR}_{\rm Ly\alpha-auto}<\textrm{SNR}_{\rm Ly\alpha\times21cm}<\textrm{SNR}_{\rm 21cm-auto}\:.
\end{equation}

As visible in Fig. \ref{fig:snrvsk_ska}, the PMF-induced enhancement in matter power leads to an increase in the SNR for all three observables at smaller scales where the effects of a PMF may be distinguished, whereas at larger scales, the SNR curves converge to a common ($\Lambda$CDM) baseline as expected. In contrast, in Fig. \ref{fig:snrvsz_ska}, the SNR for both the Ly$\alpha$ auto and the Ly$\alpha$-21 cm cross spectra are seen to peak across the intermediate redshift range $2.2\lesssim z\lesssim2.6$, falling off towards higher values of $z$ due to increasingly sparse sampling of quasars. The SNR for the 21 cm power spectrum, on the other hand, rises monotonically towards the higher end of the observable redshift range. Also, for all three observables, the rise in the SNR as a function of $k$ is steeper for sharper PMF spectra, \emph{i.e.}, more pronounced for $n_{\rm B}=-2.75$ compared to $n_{\rm B}=-2.9$, and begins at lower values of $k$. As seen in Fig. \ref{fig:snrvsk_ska}, this leads to a considerably higher SNR value for $n_{\rm B}=-2.75$ than $n_{\rm B}=-2.9$ at any given value of $k$, as increasingly smaller scales are probed. The choice of the fiducial $n_{\rm B}$ also plays an interesting role in the trend of SNR values as a function of $z$. This is reflected in Fig. \ref{fig:snrvsz_ska}, where the sequence for the different curves corresponding to distinct choices of $k$ and $B_0$ changes as $n_{\rm B}$ is varied from $n_{\rm B}=-2.9$ to $n_{\rm B}=-2.75$, in the case of the Ly$\alpha$ auto- and cross-spectra. In comparison, the sequence of the SNR curves for the 21 cm auto-spectrum remains largely unchanged.

\begin{figure*}[htbp]
    \centering
    \includegraphics[width=1.0\textwidth]{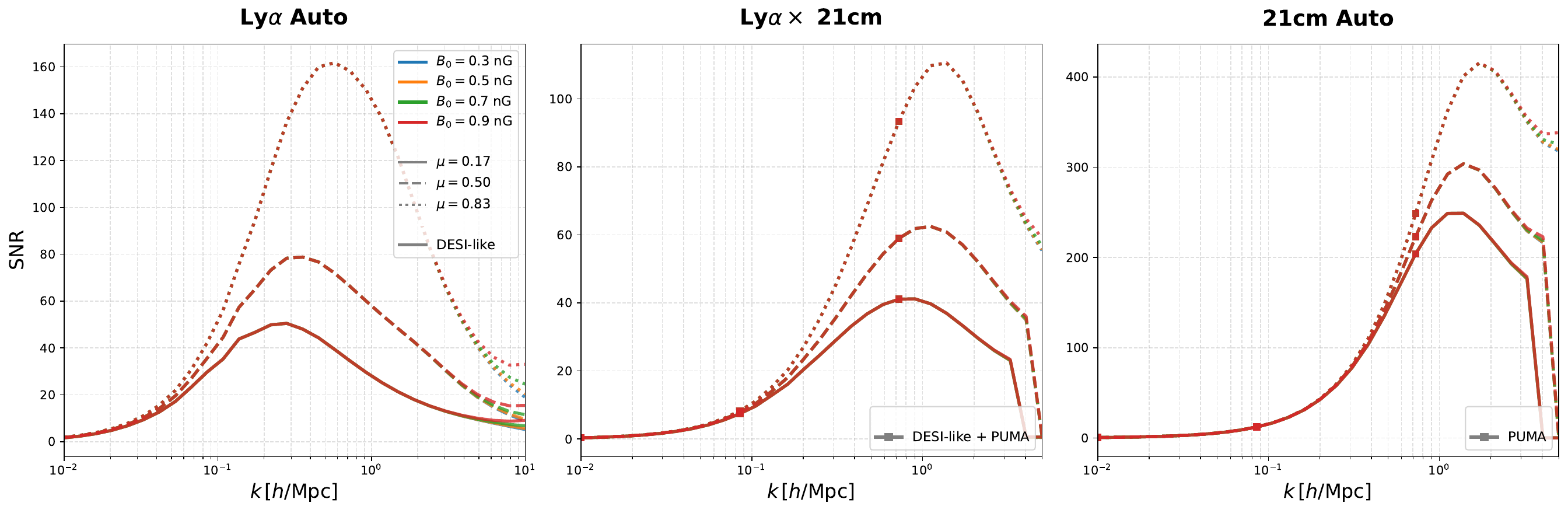}
    \includegraphics[width=1.0\textwidth]{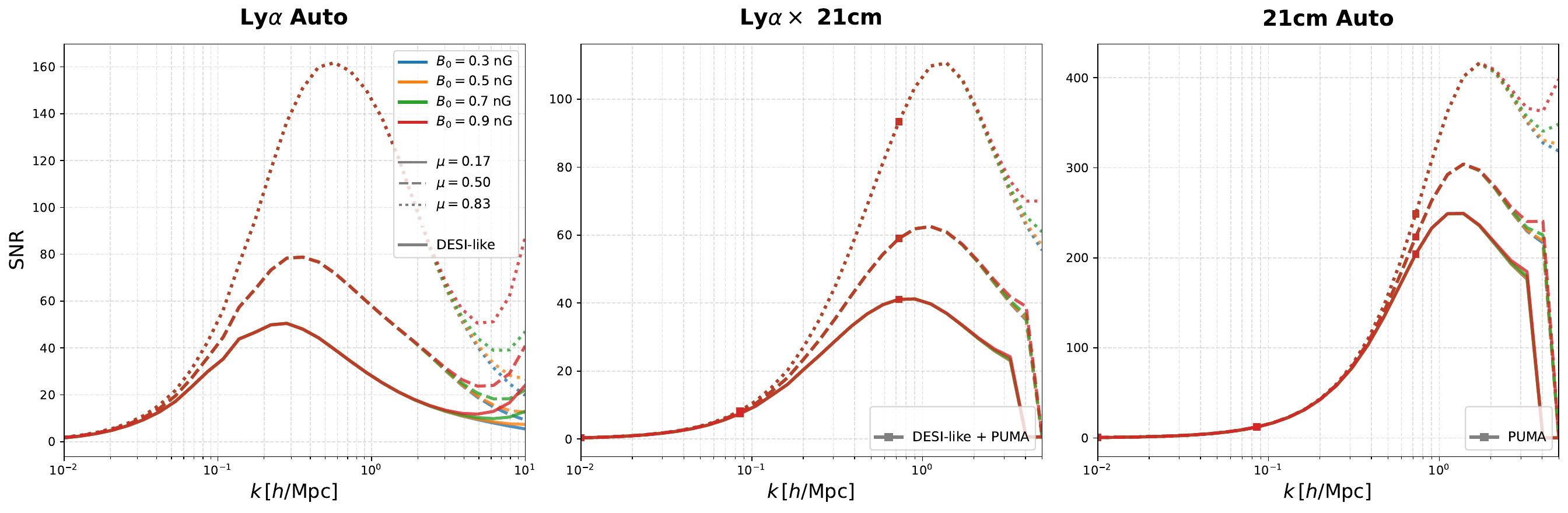}
    \caption{Signal-to-noise ratio (SNR) for the Ly$\alpha$ auto, the 21 cm auto, and the Ly$\alpha$-21 cm cross spectra shown as a function of scale, in presence of a PMF with $n_B=-2.9$ (\emph{upper panel}) and $n_{\rm B}=-2.75$ (\emph{lower panel}) and a few benchmark values of $B_0\lesssim1$ nG as mentioned in the plots. The results are shown at a fixed redshift $z=2.33$ for a few fixed values of $\mu$. For the instruments, we assume the DESI-like spectroscopic detector (for Ly$\alpha$) and the PUMA configuration (for 21 cm), as described in Sec. \ref{subsubsec:pumamodel}.}
    \label{fig:snrvsk_puma}
\end{figure*}

In Fig. \ref{fig:snrvsk_puma}, the SNR is shown similarly as a function of $k$ for all three observables with the DESI-like and PUMA combination. In this case, the maximum wavenumber for the 21 cm auto and Ly$\alpha$-21 cm cross-spectra has been set at $k=5.0\:h/\textrm{Mpc}$ due to the smaller baseline diameter of PUMA (see Table \ref{tab:puma_specs}), while the upper redshift cutoff has been fixed at $z=4$. Note that the SNR for the Ly$\alpha$ power spectrum remains identical to that in the earlier Fig. \ref{fig:snrvsk_ska}, and has been shown again due to easier comparability among the three observables for the given instrumental combination. The restriction to smaller values of $k$ for PUMA prevents the SNR from achieving very high values for distinctive values of $B_0$ in the higher $k$ regime, unlike for SKA1-Mid which may probe up to $k\sim10\:h/\textrm{Mpc}$ and is hence capable of detecting the signal where the effect of the PMF is larger. Thus, up to $k\sim5.0\:h/\textrm{Mpc}$, the projected SNR for the DESI-like+PUMA combination remains overall smaller than DESI-like+SKA1-Mid for similar fiducial values of the PMF parameters, with the 21 cm auto- and Ly$\alpha$-21 cm cross-spectra primarily tracing the $\Lambda$CDM baseline. On the other hand, we do not separately show the variation of the SNR with $z$ for fixed allowed values of $k$, as the curves virtually overlap with each other for different choices of the PMF parameters. However, for steeper PMF spectra and higher values of $B_0$, when observational data across all possible scales, redshifts, and sky directions are taken into account, PUMA might nevertheless show some constraining potential, as we explore in the next section.

Before wrapping this section up, it is imperative to make a few comments with regard to what may be expected in the case of a more realistic observational scenario. While Figs. \ref{fig:snrvsk_ska}$-$\ref{fig:snrvsk_puma} apparently demonstrate the possibility to achieve peak SNR values typically of the order $10^2-10^3$ across the three observables through the synergy of next-generation missions, it is important to note that our forecast analysis is essentially optimistic as it neglects the presence of foreground contaminants for these signals. In particular, the 21 cm auto-spectrum is expected to be severely affected by atmospheric and extraterrestrial foregrounds, which may reduce the optimistic SNR by several orders of magnitude \cite{Wolz:2015sqa,Zuo:2022wra,2022eas..conf.2230S,Diao:2024dmu}. On the other hand, while unimpeded by foregrounds, the SNR for the Ly$\alpha$ auto-spectrum is significantly lower \emph{per se}, compared to that of both the 21 cm auto-spectrum and the cross-spectrum. Interestingly, the Ly$\alpha$-21 cm cross-correlation, which exhibits an intermediate level of SNR, has the advantage of being much less impacted by such foregrounds \cite{Carucci:2016yzq}. This could make the projected SNR values for the cross-spectrum considerably more robust, thus making it a unique and efficient late-time probe that could shed light on the PMF sector.

\section{Error estimation by Fisher forecast analysis} \label{sec:fisher}

Having estimated the SNR for the three observables at the upcoming instruments, we now turn towards an analysis of the efficiency of these missions in constraining the PMF parameter space. To that end, we employ the Fisher matrix formalism \cite{Tegmark:1996bz,2009arXiv0906.0664H,Bharadwaj:2015vwa} to estimate the uncertainties in measurements of the relevant parameters by the missions under consideration. Formally, the Fisher information matrix is defined as the expectation of the negative Hessian of the log-likelihood function. In the present scenario, for a set of $n$ fiducial parameters encoded in the parameter vector $\boldsymbol{\Theta}$, the $(n\times n)$-dimensional Fisher matrix corresponding to the $i$-th observable can be expressed as
\begin{equation}
    F_{\alpha\beta}^{(i)}(\boldsymbol{\Theta})=\sum\limits_{m}^{\rm k-bins}\sum\limits_{n}^{\rm z-bins}\sum_{p}^{\rm \mu-bins}\:\dfrac{1}{\sigma^2\left[P_i\left(k_m,z_n,\mu_p\right)\right]}\dfrac{\partial P_i\left(k_m,z_n,\mu_p\right)}{\partial\Theta_\alpha}\dfrac{\partial P_i\left(k_m,z_n,\mu_p\right)}{\partial\Theta_\beta}\:,
\end{equation}
where $\alpha$ and $\beta$ run from $1$ to $n$. The inverse of the Fisher matrix then gives the covariance matrix, whose diagonal elements correspond to the projected $1\sigma$ errors for the parameters, and off-diagonal elements correspond to the correlation among different parameters. In case of a degeneracy between two or more parameters, the Fisher matrix is singular, which reflects the inability of the given dataset to distinguish between, and hence individually constrain, the degenerate parameters.

\begin{figure*}[htbp]
    \centering
    \includegraphics[width=0.48\textwidth]{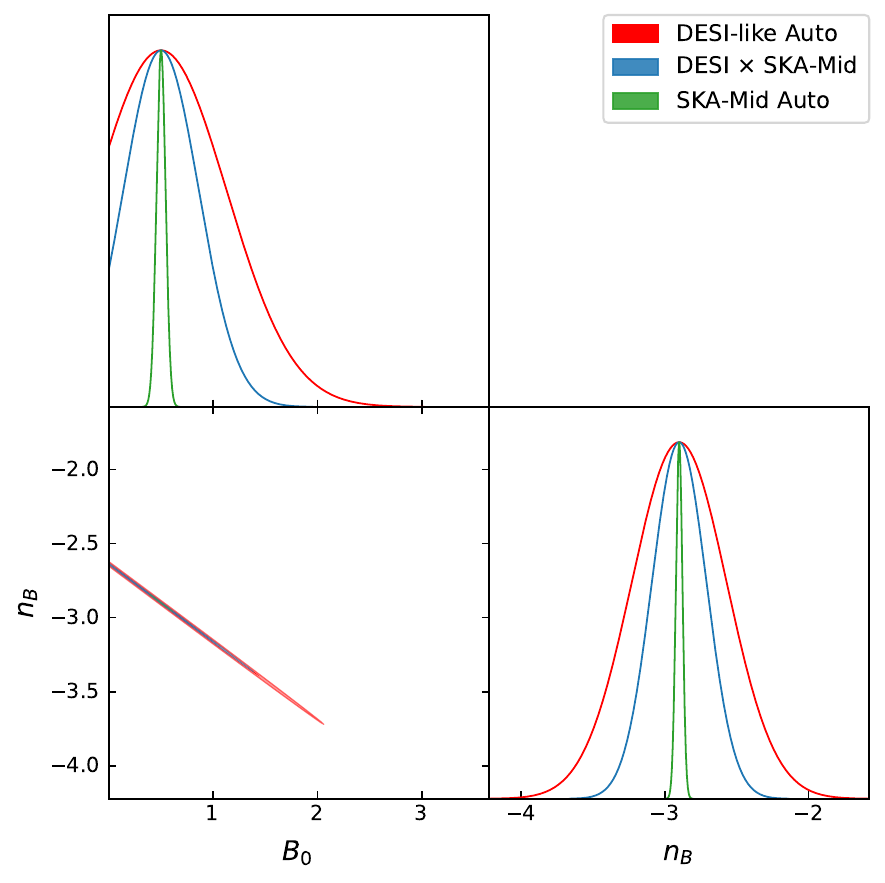}
    \includegraphics[width=0.48\textwidth]{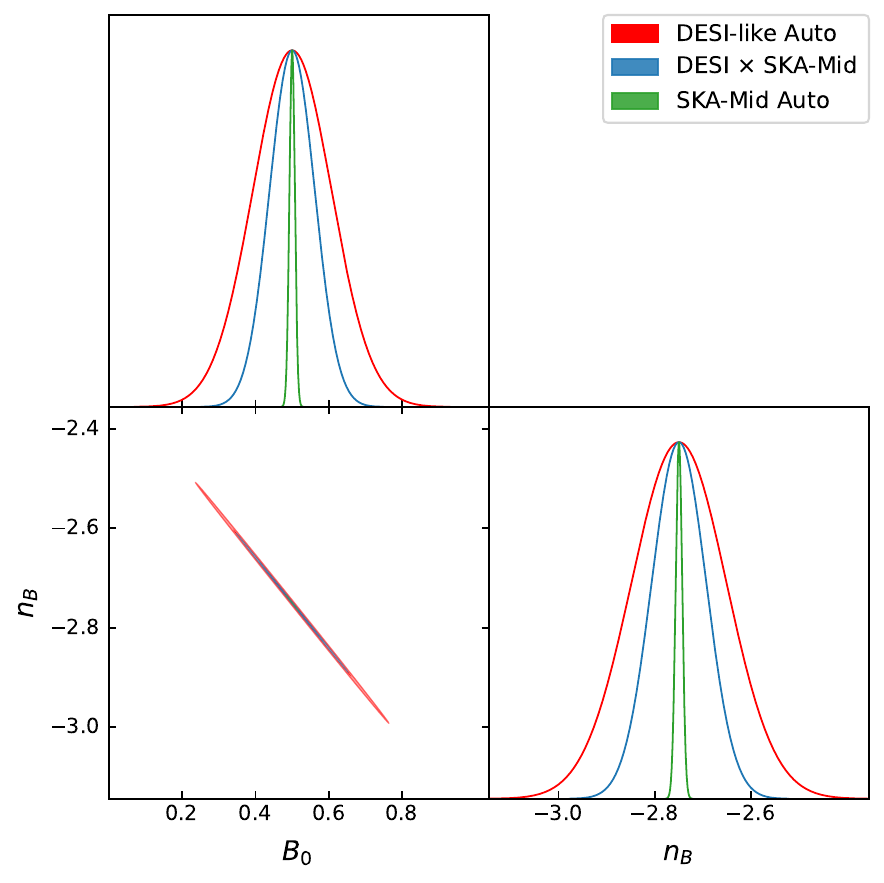}
    \includegraphics[width=0.48\textwidth]{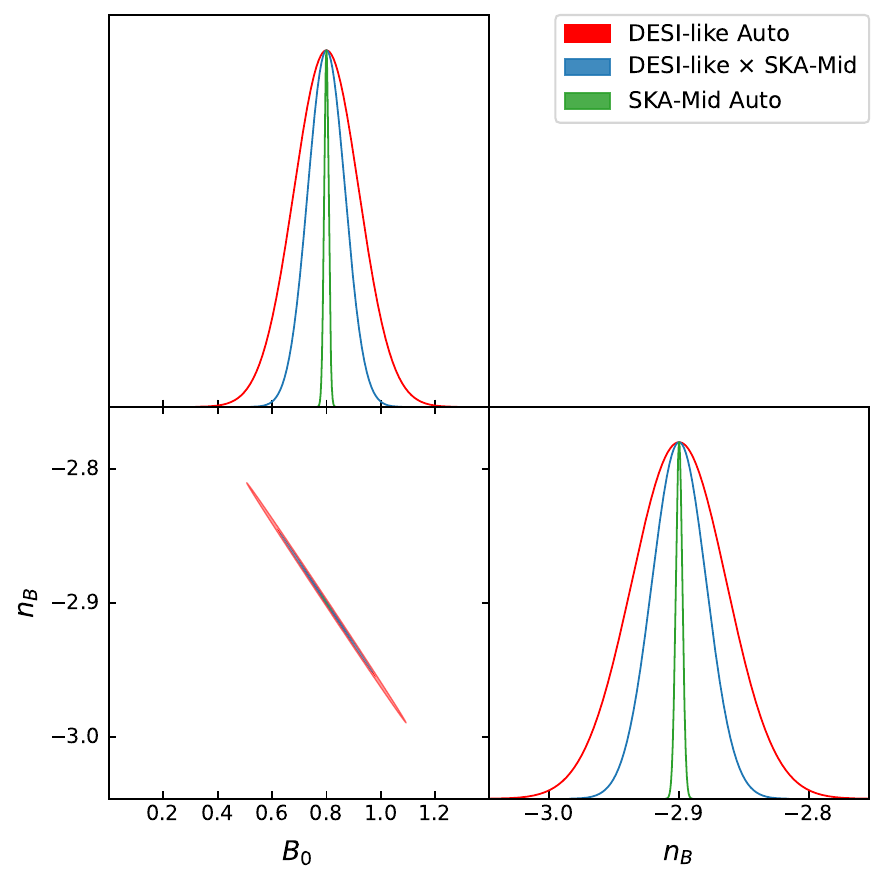}
    \includegraphics[width=0.48\textwidth]{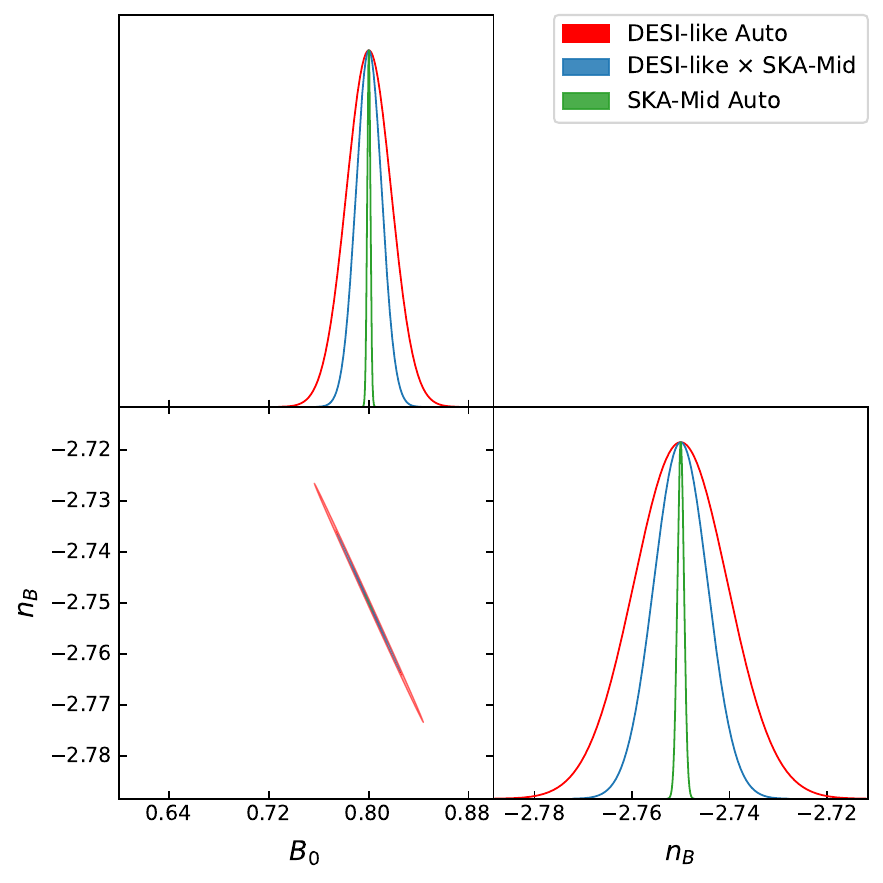}
    \caption{Two-dimensional contour plots in the $B_0-n_{\rm B}$ plane showing projected 1$\sigma$ errors and correlation of the PMF parameters, based on each of the three different observable signals with the DESI-like and SKA1-Mid instrumental combination. The fiducial values chosen for the plot are $B_0=0.5$ nG (\emph{first row}) and $B_0=0.8$ nG (\emph{second row}), and $n_{\rm B}=-2.9$ (\emph{first column}) and $n_{\rm B}=-2.75$ (\emph{second column}). }
    \label{fig:fisher2d_ska}
\end{figure*}

The Fisher posterior distributions and correlation ellipses for the two PMF parameters $B_0$ and $n_{\rm B}$ are shown in Fig. \ref{fig:fisher2d_ska} for all three signals based on the DESI-like+SKA1-Mid combination, corresponding to two sets of representative fiducial values $B_0=\{0.5\:\textrm{nG},\:0.8\:\textrm{nG}\}$ and $n_{\rm B}=\{-2.9,-2.75\}$. For our analysis, we have taken $10$ logarithmically equispaced $k$-bins in $k\in\left[0.1,10\right]\:h/\textrm{Mpc}$, $5$ equispaced $z$-bins in $z\in\left[2.00,3.05\right]$, and $5$ equispaced $\mu$-bins in $\mu\in\left[0,1\right]$. Let us now turn to the discussion of a few important features that immediately become apparent from Fig. \ref{fig:fisher2d_ska}. Firstly, the constraining power of the three different observables is found to follow their overall SNR trend observed in Sec. \ref{sec:snr}, \emph{i.e.}, for a given set of fiducial values of the PMF parameters, the 21 cm power spectrum is seen to offer the tightest bounds, followed by the Ly$\alpha$-21 cm cross-spectrum and finally the Ly$\alpha$ power spectrum. For example, based on the synergy of DESI-like+SKA1-Mid, our analysis indicates that the amplitude of a PMF of mean strength $B_0=0.8$ nG and spectral tilt $n_{\rm B}=-2.9$ may be constrained with a $1\sigma$ error bound of $\Delta B_0\approx0.12$ nG via the Ly$\alpha$ auto-spectrum, $\Delta B_0\approx0.07$ nG via the Ly$\alpha$-21 cm cross-spectrum, and $\Delta B_0\approx0.01$ nG via the 21-cm auto-spectrum. Secondly, the projected $1\sigma$ errors on both $B_0$ and $n_{\rm B}$ diminish for a higher value of $B_0$ and a steeper value of $n_{\rm B}$, which is readily apparent upon comparing both row and column-wise among the different triangle plots in Fig. \ref{fig:fisher2d_ska}. This is expected due to greater enhancement of the underlying matter power for stronger PMF amplitudes and sharper PMF spectra (see Fig. \ref{fig:magnetic_panels}), which is also reflected in the SNR results of Sec. \ref{sec:snr}. Importantly, we find that even a slight variation in the value of the PMF spectral index leads to drastic changes in the projected error values. For example, for the fiducial value of $B_0=0.5$ nG, the Ly$\alpha$ power spectrum and the Ly$\alpha$-21 cm cross-spectrum seem capable of providing only very loose upper bounds on $B_0$ for the nearly scale-invariant case $n_{\rm B}=-2.9$. On the other hand, for a slightly sharper PMF spectrum with $n_{\rm B}=-2.75$, the same two observables end up furnishing simultaneous upper and lower limits on $B_0$ with a relative $1\sigma$ error bound $\Delta B_0/B_0\sim10\%$. Similarly, the relative $1\sigma$ bound for the fiducial value $B_0=0.8$ nG shows nearly one full order of magnitude improvement across all three observables upon switching the fiducial value of $n_{\rm B}$ from $n_{\rm B}=-2.9$ to $n_{\rm B}=-2.75$. Finally, all three observables apparently display a sharp negative correlation between the parameters $B_0$ and $n_{\rm B}$, which is quantified by the narrow spread of the Fisher ellipse. 

\begin{figure*}[htbp]
    \centering
    \includegraphics[width=0.48\textwidth]{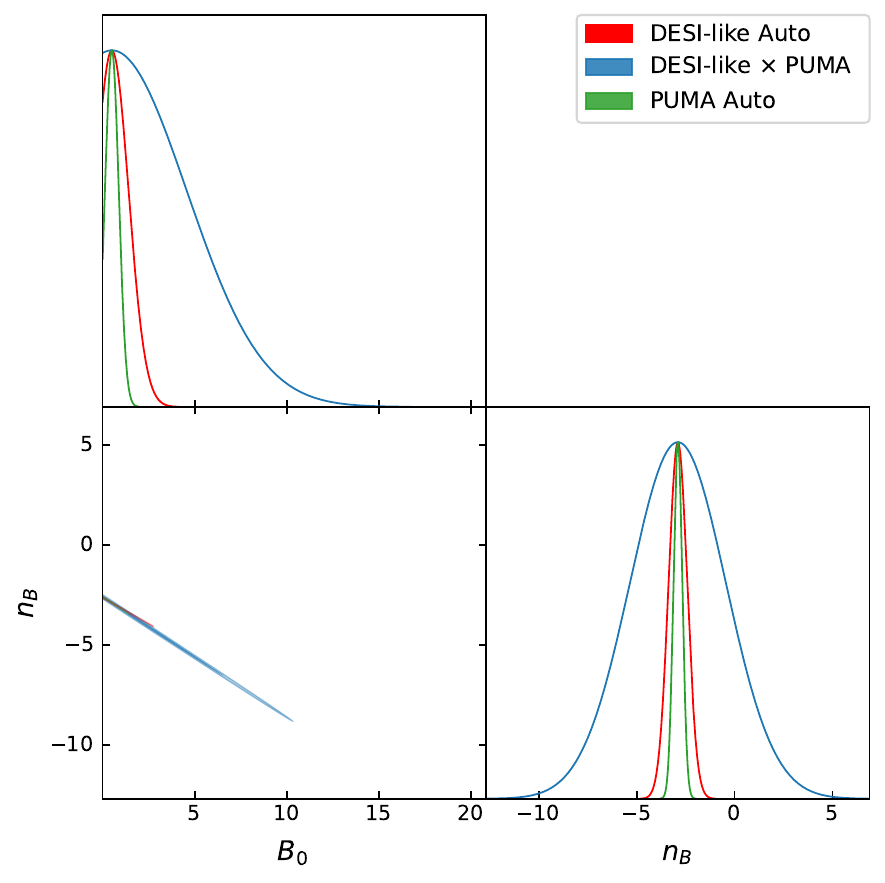}
    \includegraphics[width=0.48\textwidth]{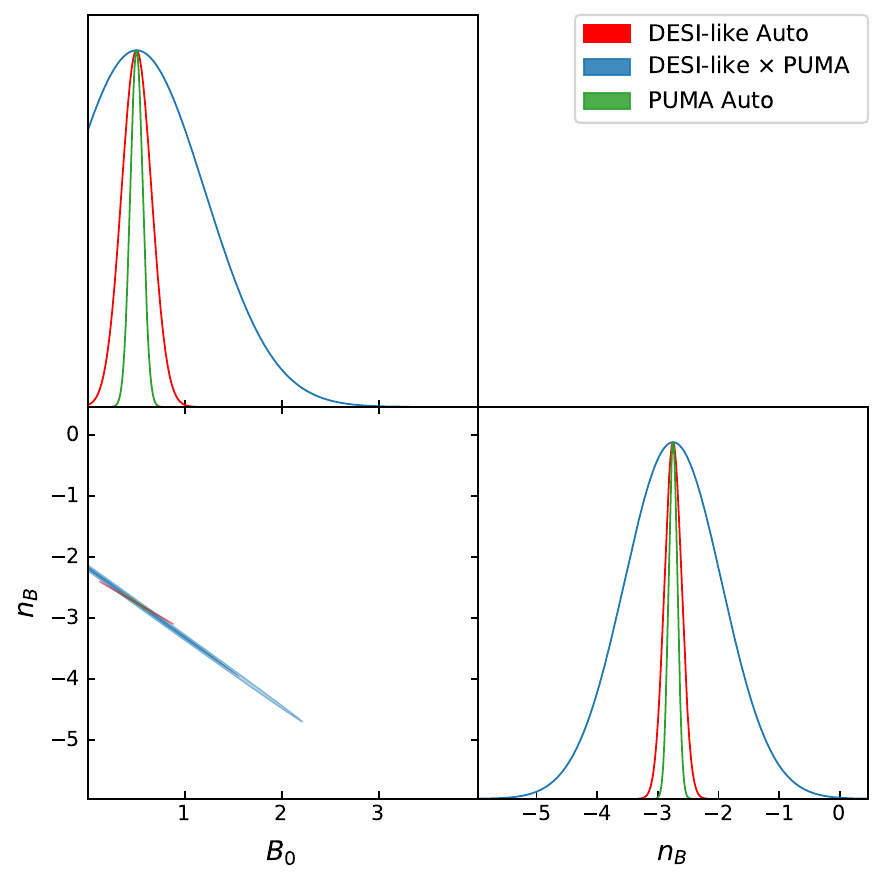}
    \includegraphics[width=0.48\textwidth]{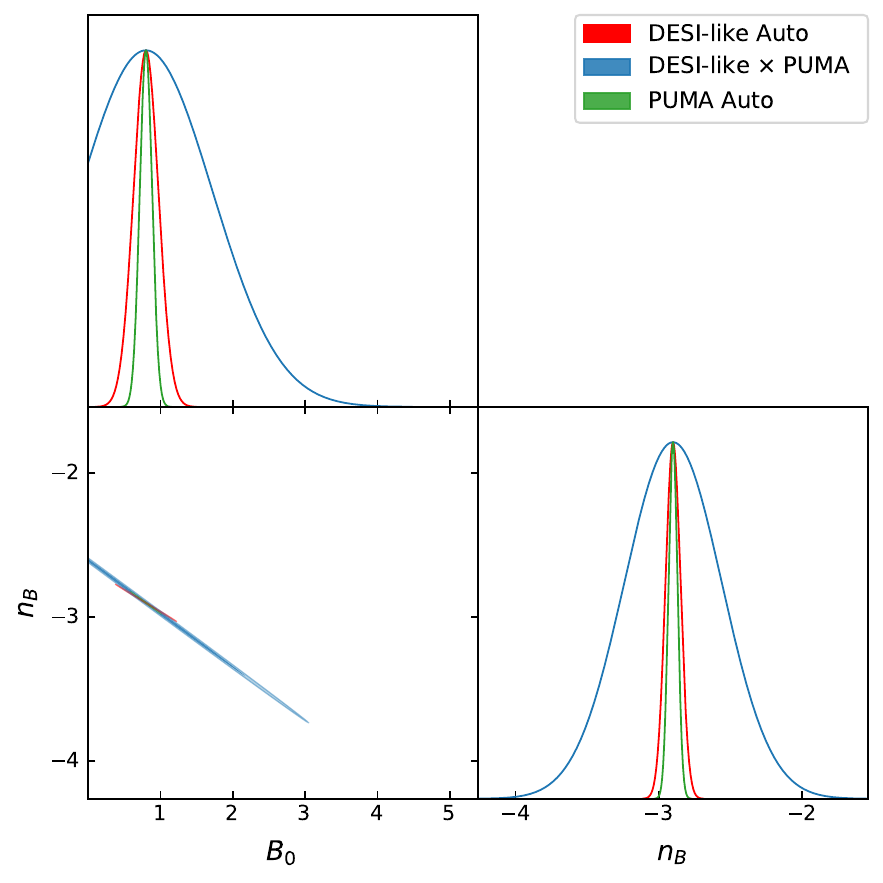}
    \includegraphics[width=0.48\textwidth]{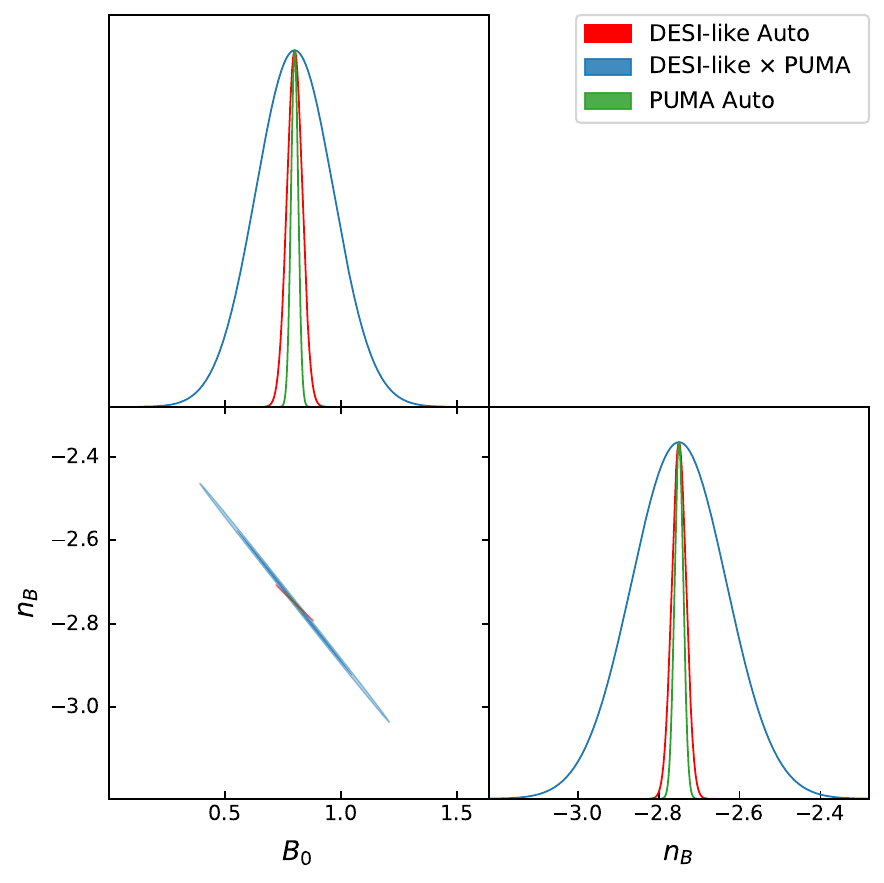}
    \caption{Two-dimensional contour plots in the $B_0-n_{\rm B}$ plane showing projected 1$\sigma$ errors and correlation of the PMF parameters, based on each of the three different observable signals with the DESI-like and PUMA instrumental combination. The fiducial values chosen for the plot are $B_0=0.5$ nG (\emph{first row}) and $B_0=0.8$ nG (\emph{second row}), and $n_{\rm B}=-2.9$ (\emph{first column}) and $n_{\rm B}=-2.75$ (\emph{second column}). }
    \label{fig:fisher2d_puma}
\end{figure*}

In Fig. \ref{fig:fisher2d_puma}, the same analysis has been repeated for the DESI-like+PUMA combination. The number of bins used is the same as before, with the only difference being that the $k$ and $z$ ranges have now been fixed at $k\in\left[0.1,5.0\right]\:h/\textrm{Mpc}$ and $z\in\left[2,4\right]$. While PUMA is expected to probe larger scales with much better precision than SKA1-Mid due to a considerably higher number of dishes, we actually observe consistently superior performance by SKA1-Mid compared to PUMA in the case of the PMF-enhanced matter sector considered in this study. This is chiefly because the shorter baseline of PUMA prevents it from probing a significant portion of the $k$-range where the matter power enhancement induced by the PMF is discernibly large. As a result, the 21 cm sector loses constraining power due to the reduced range of scales, which manifests as wider $1\sigma$ bounds furnished by the 21 cm auto-spectrum and the Ly$\alpha$-21 cm cross-spectrum. For instance, based on the synergy of DESI-like+PUMA, our analysis reveals that the amplitude of a PMF of mean strength $B_0=0.8$ nG and spectral tilt $n_{\rm B}=-2.75$ could be constrained with a $1\sigma$ error bound of $\Delta B_0\approx0.03$ nG via the Ly$\alpha$ auto-spectrum, $\Delta B_0\approx0.17$ nG via the Ly$\alpha$-21 cm cross-spectrum, and $\Delta B_0\approx0.02$ nG via the 21-cm auto-spectrum. 
This is revealed through a direct comparison between Figs. \ref{fig:fisher2d_ska} and \ref{fig:fisher2d_puma}, which indicates a consistently superior performance of the DESI-like+SKA1-Mid combination compared to DESI-like+PUMA in constraining the parameters describing a sub-nG stochastic PMF via the cross-correlation signal. In fact, as visible in Fig. \ref{fig:fisher2d_puma}, the cross-spectrum provides the weakest bounds among all three observables. On the other hand, since the maximum redshift has been fixed at $z=4$ for compatibility with PUMA, a uniform $z$-binning with $5$ bins (which increases the bin width) slightly worsens the performance of the DESI-like instrument compared to the synergistic case with SKA1-Mid, since Ly$\alpha$ QSO samples become sparser at higher redshifts. Therefore, the constraining potential of the Ly$\alpha$ auto- and cross-spectra is somewhat compromised. Arguably, this could be ameliorated by keeping the bin width the same as that for DESI-like+SKA1-Mid and considering a higher number of $z$-bins to extend across the larger redshift range, which would then have resulted instead in slightly tighter bounds for the Ly$\alpha$ auto-spectrum. However, our binning choice is motivated by the requirement of keeping a constant number of bins, which is more instructive from the perspective of comparing between the synergy of both combinations. 
Consequently, the only fiducial pair for which the Ly$\alpha$ auto-spectrum results in a relative error on $B_0$ marginally less than unity is $B_0=0.8$ nG and $n_{\rm B}=-2.75$, which, for the DESI-like+SKA1-Mid combination, results in a relative error of around $10\%$. In Table \ref{tab:B0_nB_error}, we have listed the numerical values of the projected $1\sigma$ bounds on the PMF parameters that appear in Figs. \ref{fig:fisher2d_ska} and \ref{fig:fisher2d_puma}, as functions of their fiducial values for all combinations of observables and missions under consideration.

\begin{figure}[htbp]
    \centering
    \includegraphics[width=0.48\linewidth]{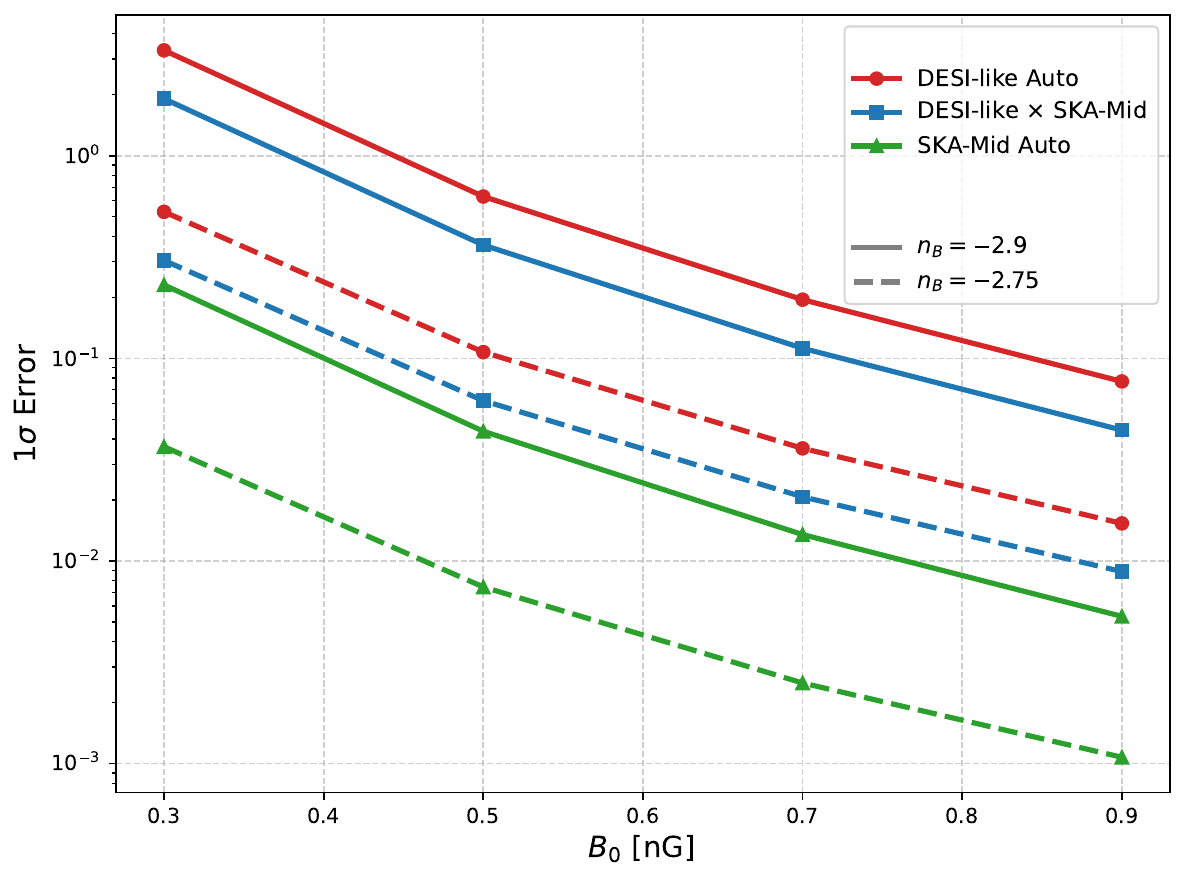}
    \includegraphics[width=0.48\linewidth]{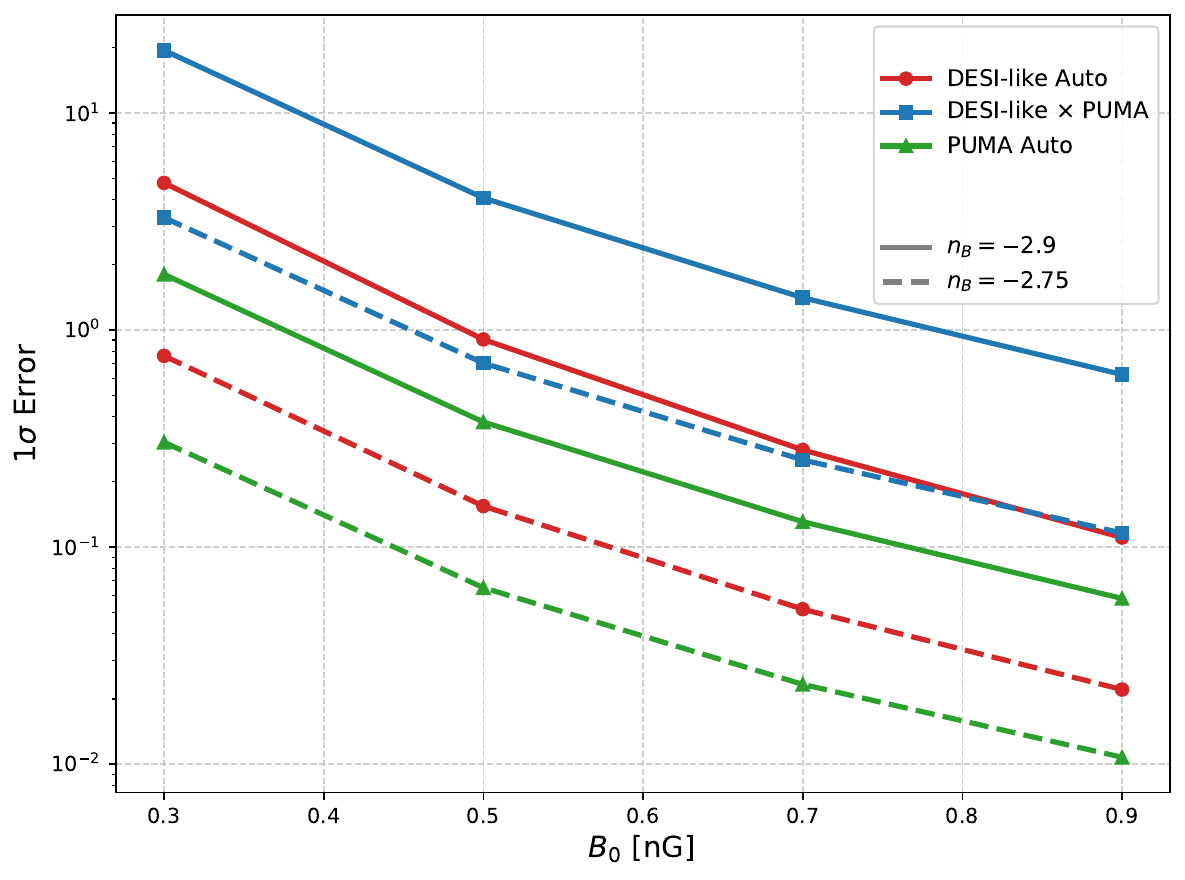}
    \caption{Dependence of the projected (absolute) 1$\sigma$ error of $B_0$ on the fiducial value of $B_0$ for $n_{\rm B}=-2.9$ (\emph{solid lines}) and $n_{\rm B}=-2.75$ (\emph{dashed lines}), based on the three observable signals with the DESI-like+SKA1-Mid (\emph{left}) and DESI-like+PUMA (\emph{right}) combinations.}
    \label{fig:fiducial_wrt_B0}
\end{figure}

While Figs. \ref{fig:fisher2d_ska} and \ref{fig:fisher2d_puma} already hint at some of the key dependencies on fiducial choices and shed light on the relative efficacy of the two different instrumental combinations in constraining the PMF sector, it is nonetheless prudent to consider a wider fiducial range that should make the general trends more transparent and therefore allow one to arrive at more robust conclusions. For a clearer comparison between the performances of the DESI-like+SKA1-Mid and DESI-like+PUMA configurations, the dependence of the predicted $1\sigma$ bound on $B_0$ upon the fiducial values of $B_0$ and $n_{\rm B}$ has been plotted in Fig. \ref{fig:fiducial_wrt_B0} for all three observables, corresponding separately to each instrumental combination. The plots reflect the general trend of the fiducial dependence of the forecast $1\sigma$ error, $\Delta B_0$, across the range $0.1\:\textrm{nG}<B_0<1.0\:\textrm{nG}$, in the nearly scale-invariant regime with $n_{\rm B}=-2.9$ \emph{vis-\`{a}-vis} the slightly sharper case with $n_{\rm B}=-2.75$. For both DESI-like+SKA1-Mid and DESI-like+PUMA, the 21 cm auto-spectrum apparently shows the strongest constraining power. With SKA1-Mid, the 21 cm auto-spectrum is seen to be capable of furnishing $\Delta B_0/B_0\lesssim10\%$ for $B_0\gtrsim0.5$ nG in the case of $n_{\rm B}=-2.9$, and for $B_0\gtrsim0.3$ nG in the case of $n_{\rm B}=-2.75$. The next best level of constraint is provided by the Ly$\alpha$-21 cm cross-spectrum in the case of DESI-like+SKA1-Mid, and the Ly$\alpha$ auto-spectrum in the case of DESI-like+PUMA, with both cases furnishing, on an average, one order of magnitude weaker bounds than the 21 cm auto-spectrum for identical choices of fiducial parameters. Specifically, in the slightly sharper case $n_{\rm B}=-2.75$, the Ly$\alpha$-21 cm cross-spectrum via DESI-like+SKA1-Mid seems capable of furnishing $\Delta B_0/B_0\lesssim10\%$ for $B_0\gtrsim0.5$ nG, whereas the Ly$\alpha$ auto-spectrum via DESI-like+PUMA could yield $\Delta B_0/B_0\lesssim10\%$ for $B_0\gtrsim0.4$ nG.  In view of the foreground contamination of the realistic 21 cm auto-spectrum, the uncontaminated post-EoR Ly$\alpha$-21 cm cross-spectrum, probed via the synergy of the next-generation DESI-like+SKA1-Mid instrumental combination, thus appears to be a significantly robust and promising channel to constrain a weakly scale-dependent sub-nG PMF based solely on its impact on small-scale matter clustering.

\begin{table}[htbp]
\centering
\begin{tabular}{ |c|c|c|c|c| } 
\hline
Configuration & Observable & Fiducial values & $n_{\rm B}=-2.9$ & $n_{\rm B}=-2.75$ \\
\hline

\multirow{12}{*}{DESI-like+SKA1-Mid}

& \multirow{4}{*}{Ly$\alpha$ auto} 
& \multirow{2}{4em}{$B_0=0.5$} & $\Delta B_0=0.62924$ & $\Delta B_0=0.10737$ \\ 
&  &  & $\Delta n_{\rm B}=0.33070$ & $\Delta n_{\rm B}=0.09855$ \\ 
\cline{3-5}
&  & \multirow{2}{4em}{$B_0=0.8$} & $\Delta B_0=0.11963$ & $\Delta B_0=0.01792$ \\ 
&  &  & $\Delta n_{\rm B}=0.03659$ & $\Delta n_{\rm B}=0.0096$ \\ 
\cline{2-5}

& \multirow{4}{*}{Ly$\alpha$-21 cm} 
& \multirow{2}{4em}{$B_0=0.5$} & $\Delta B_0=0.36252$ & $\Delta B_0=0.06182$ \\ 
&  &  & $\Delta n_{\rm B}=0.19023$ & $\Delta n_{\rm B}=0.05665$ \\ 
\cline{3-5}
&  & \multirow{2}{4em}{$B_0=0.8$} & $\Delta B_0=0.06881$ & $\Delta B_0=0.01034$ \\ 
&  &  & $\Delta n_{\rm B}=0.02099$ & $\Delta n_{\rm B}=0.00553$ \\ 
\cline{2-5}

& \multirow{4}{*}{21 cm auto} 
& \multirow{2}{4em}{$B_0=0.5$} & $\Delta B_0=0.04366$ & $\Delta B_0=0.00744$ \\ 
&  &  & $\Delta n_{\rm B}=0.022901$ & $\Delta n_{\rm B}=0.00682$ \\ 
\cline{3-5}
&  & \multirow{2}{4em}{$B_0=0.8$} & $\Delta B_0=0.0082$ & $\Delta B_0=0.00128$ \\ 
&  &  & $\Delta n_{\rm B}=0.00253$ & $\Delta n_{\rm B}=0.00068$ \\ 
\hline

\multirow{12}{*}{DESI-like+PUMA}

& \multirow{4}{*}{Ly$\alpha$ auto} 
& \multirow{2}{4em}{$B_0=0.5$} & $\Delta B_0=0.90629$ & $\Delta B_0=0.15464$ \\ 
&  &  & $\Delta n_{\rm B}=0.47630$ & $\Delta n_{\rm B}=0.14193$ \\ 
\cline{3-5}
&  & \multirow{2}{4em}{$B_0=0.8$} & $\Delta B_0=0.17229$ & $\Delta B_0=0.03306$ \\ 
&  &  & $\Delta n_{\rm B}=0.0527$ & $\Delta n_{\rm B}=0.01770$ \\ 
\cline{2-5}

& \multirow{4}{*}{Ly$\alpha$-21 cm} 
& \multirow{2}{4em}{$B_0=0.5$} & $\Delta B_0=4.06559$ & $\Delta B_0=0.70494$ \\ 
&  &  & $\Delta n_{\rm B}=2.44937$ & $\Delta n_{\rm B}=0.80301$ \\ 
\cline{3-5}
&  & \multirow{2}{4em}{$B_0=0.8$} & $\Delta B_0=0.91814$ & $\Delta B_0=0.16739$ \\ 
&  &  & $\Delta n_{\rm B}=0.34012$ & $\Delta n_{\rm B}=0.11742$ \\ 
\cline{2-5}

& \multirow{4}{*}{21 cm auto} 
& \multirow{2}{4em}{$B_0=0.5$} & $\Delta B_0=0.37747$ & $\Delta B_0=0.06506$ \\ 
&  &  & $\Delta n_{\rm B}=0.22794$ & $\Delta n_{\rm B}=0.07431$ \\ 
\cline{3-5}
&  & \multirow{2}{4em}{$B_0=0.8$} & $\Delta B_0=0.08531$ & $\Delta B_0=0.01548$ \\ 
&  &  & $\Delta n_{\rm B}=0.03169$ & $\Delta n_{\rm B}=0.0109$ \\ 
\hline

\end{tabular}
\caption{The $1\sigma$ Fisher error forecasts for different fiducial combinations of $B_0$ and $n_B$, for all three observables across two different instrumental combinations. Note that the units of $B_0$ and $\Delta B_0$ are in nG.}
\label{tab:B0_nB_error}
\end{table}

Having focused so far only on the two PMF parameters, we now proceed to perform a joint Fisher analysis combining the six $\Lambda$CDM parameters and the two PMF parameters, based on all three of the post-EoR observables for both DESI-like+SKA1-Mid and DESI-like+PUMA. Empowered with the observational synergy offered by next-generation missions, this approach enables us to assess the standalone potential of each post-EoR observable in constraining the effective $(6+2)-$parameter cosmological model that results from augmenting the vanilla $\Lambda$CDM scenario with a non-helical stochastic PMF sector. The $\Lambda$CDM parameter space is spanned by the following six parameters: the reduced baryon density parameter ($\omega_b$), the reduced CDM density parameter ($\omega_{\rm cdm}$), the reduced dimensionless Hubble's constant ($h$), the scalar power amplitude ($A_s$) and the scalar spectral index ($n_s$) at the CMB pivot scale $k_*=0.05\:\textrm{Mpc}^{-1}$, and the optical depth to reionization ($\tau$). In the subsequent analysis, the fiducial values chosen for these six parameters are their respective best fit-values reported by \textit{Planck} 2018 \cite{Planck:2018vyg}. On the other hand, the PMF parameter space consists of the two parameters $B_0$ and $n_{\rm B}$ as before, whose fiducial values have been fixed at $B_0=0.8$ nG and $n_{\rm B}=-2.75$ for the purpose of illustration. 

\begin{figure}[htbp]
    \centering
    \includegraphics[width=1.0\linewidth]{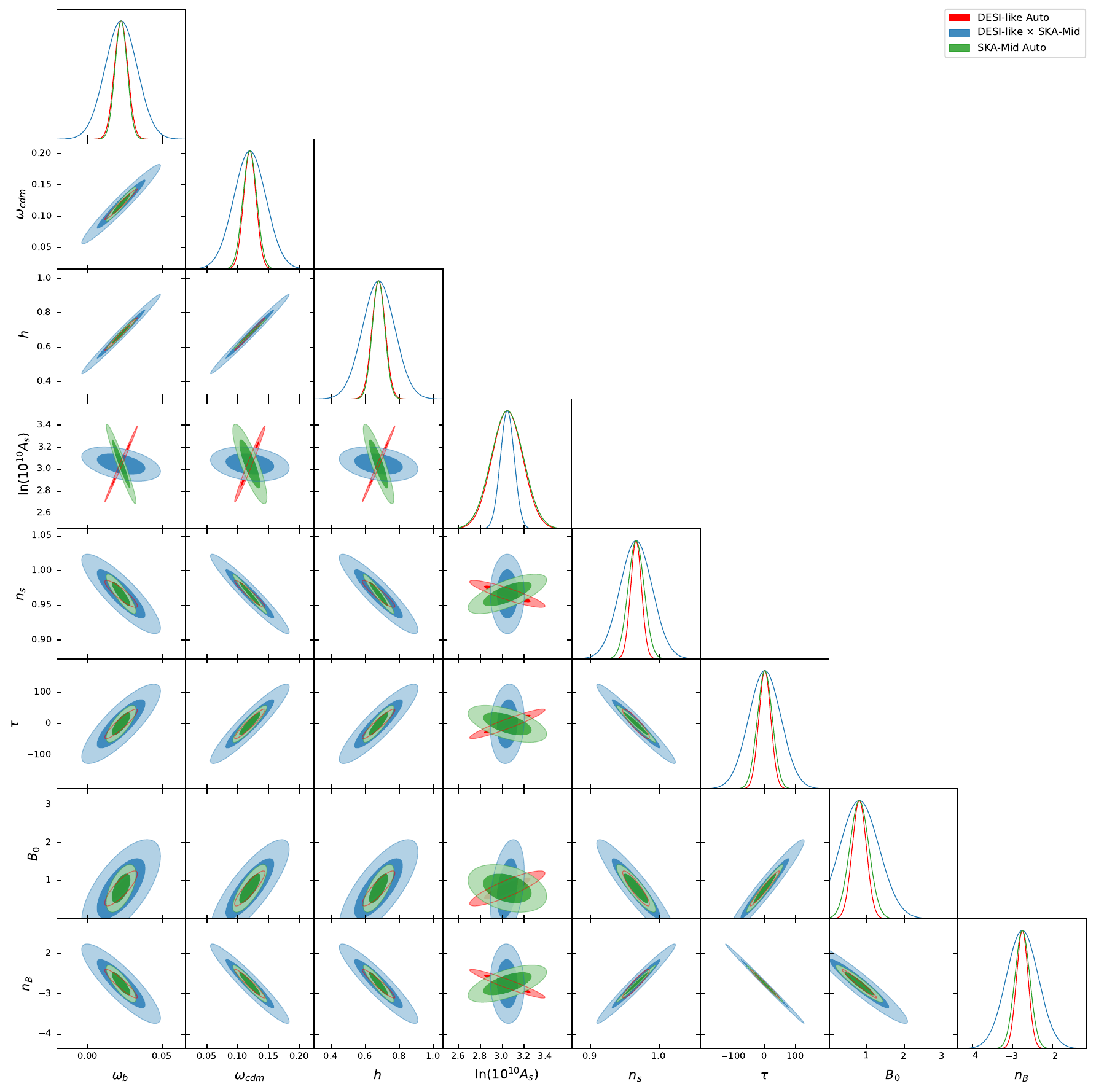}
    \caption{Fisher triangle plots for the $6+2$ parameters spanning the joint $\Lambda$CDM+PMF parameter space based on the following three post-EoR observables: Ly$\alpha$ auto-spectrum (probed via DESI-like), the Ly$\alpha$-21 cm cross-spectrum (probed via DESI-like+SKA1-Mid), and the 21 cm auto-spectrum (probed via SKA1-Mid). The darker and lighter shades correspond to projected $1\sigma$ and $2\sigma$ contours respectively. The $\Lambda$CDM fiducial values have been taken to be the best-fit values reported by \textit{Planck} 2018 \cite{Planck:2018vyg}. The fiducial values of the PMF parameters have been fixed at $B_0=0.8$ nG and $n_{\rm B}=-2.75$.}
    \label{fig:fisher_8param_SKA}
\end{figure}

\begin{figure}[htbp]
    \centering
    \includegraphics[width=1.0\linewidth]{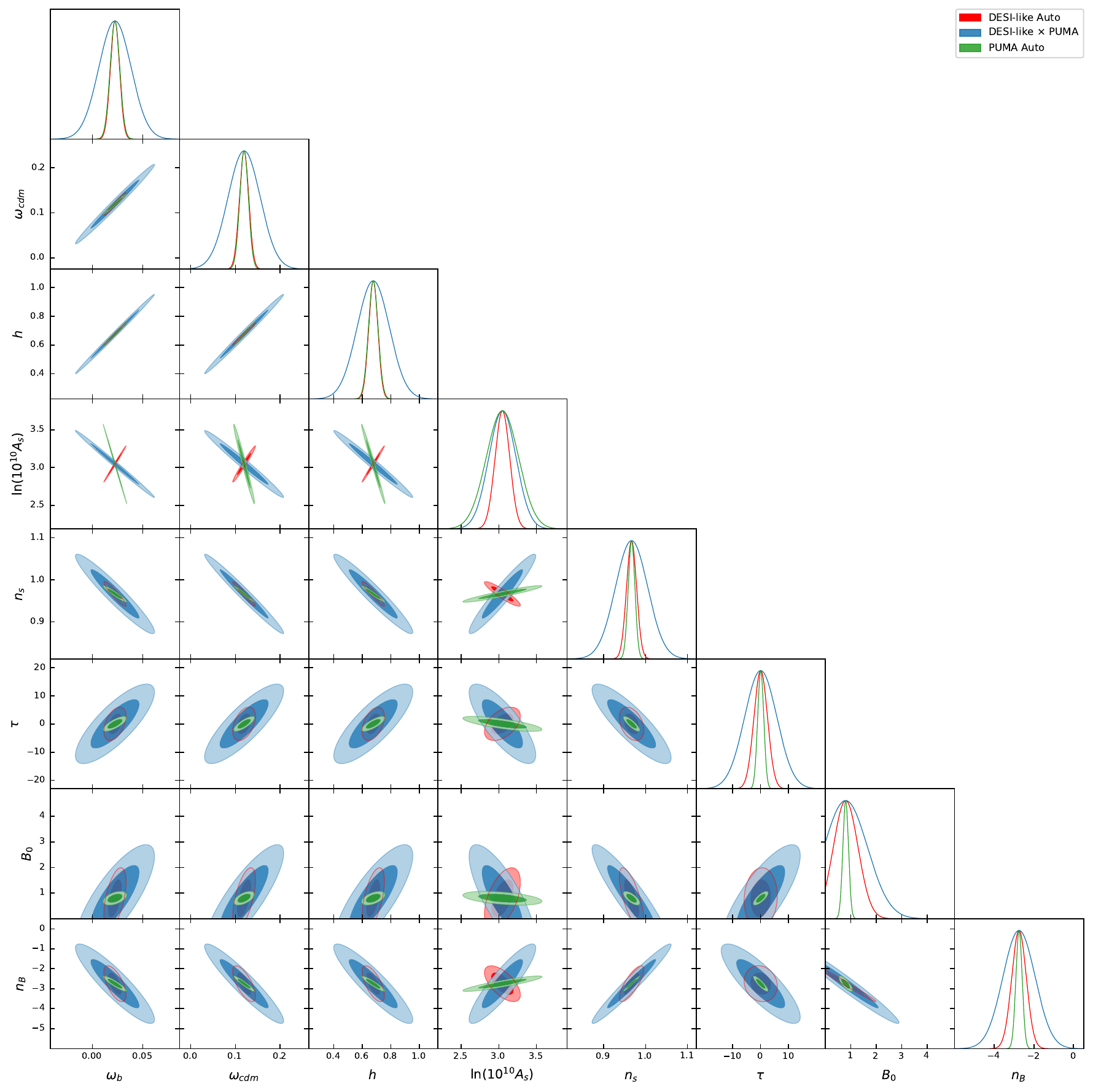}
    \caption{Fisher triangle plots for the $6+2$ parameters spanning the joint $\Lambda$CDM+PMF parameter space based on the following three post-EoR observables: Ly$\alpha$ auto-spectrum (probed via DESI-like), the Ly$\alpha$-21 cm cross-spectrum (probed via DESI-like+PUMA), and the 21 cm auto-spectrum (probed via PUMA). The darker and lighter shades correspond to projected $1\sigma$ and $2\sigma$ contours respectively. The $\Lambda$CDM fiducial values have been taken to be the best-fit values reported by \textit{Planck} 2018 \cite{Planck:2018vyg}. The fiducial values of the PMF parameters have been fixed at $B_0=0.8$ nG and $n_{\rm B}=-2.75$.}
    \label{fig:fisher_8param_PUMA}
\end{figure}

Fig. \ref{fig:fisher_8param_SKA} shows the projected Fisher posterior distributions for each of the $8$ individual parameters as well as the $1\sigma$ and $2\sigma$ correlation ellipses for each possible pair, based separately on each of the three observables and corresponding to the DESI-like+SKA1-Mid instrumental combination. In Fig. \ref{fig:fisher_8param_PUMA}, the same analysis has been repeated for the DESI-like+PUMA combination. As expected, the inclusion of the six $\Lambda$CDM parameters considerably degrades the constraining power on the PMF sector in each case, as revealed by a direct comparison with the lower right triangle plots in both Figs. \ref{fig:fisher2d_ska} and \ref{fig:fisher2d_puma}. This visibly manifests as wider correlation contours and larger projected errors on $B_0$ and $n_{\rm B}$ for all three of the observables. Additionally, for DESI-like+PUMA, simultaneous analysis across $(6+2)-$parameters indicates a similar relative trend in constraining power for the three observables as compared to the case with only $2$ PMF parameters in Fig. \ref{fig:fisher2d_puma}. On the other hand, compared to Fig. \ref{fig:fisher2d_ska}, the trend noticeably changes in the case of DESI-like+SKA1-Mid once the full parameter space is taken into account, with the Ly$\alpha$ auto-spectrum predicted to provide comparatively better constraints than the Ly$\alpha$-21 cross-spectrum on all $6+2$ parameters.

In a realistic parameter estimation scenario, post-EoR observables alone are anyhow not expected to yield tight constraints on the six parameters of the $\Lambda$CDM model. For substantial improvement in the error bounds on these parameters, CMB TT+TE+EE datasets need to be taken into account. However, these datasets are not expected to lead to significantly tight levels of constraint on a nearly scale-invariant sub-nG PMF, whose effects on these spectra are anticipated to be sub-dominant compared to the $\Lambda$CDM baseline \cite{Zucca:2016iur}. On the other hand, the CMB BB spectrum is expected to bear a much clearer imprint of such a PMF, but simultaneously render it strongly degenerate with the signature of inflationary gravitational waves \cite{Zucca:2016iur}. Thus, one may expect a separation of roles as far as the two different kinds of datasets are concerned, with current and future CMB temperature and $E$-mode polarization data providing strong bounds on the six parameters of the $\Lambda$CDM scenario, and post-EoR observables playing a distinctive role in constraining the PMF sector. In spirit, such a scenario then appears to be equivalent to our earlier approach, which consists of drawing the baseline $\Lambda$CDM fiducial priors from existing CMB constraints and forecasting on the PMF sector based on mock post-EoR data generated using those fiducials.

\section{Summary and future directions} \label{sec:concl}

We have investigated the impact of a stochastic, Gaussian, non-helical PMF on three post-EoR cosmological observables, \emph{i.e.}, the Ly$\alpha$ power spectrum, the 21 cm power spectrum, and the Ly$\alpha$-21 cm cross-spectrum, and assessed the capability of a few prominent next-generation cosmic surveys in constraining such a nearly scale-invariant PMF of sub-nG strength based on the possible detection of these three signals. Such a PMF is characterized by its present-day comoving amplitude smoothed on a comoving scale of $1\:\textrm{Mpc}$, $B_0$, and the magnetic spectral index characterizing the slope of the PMF power spectrum, $n_{\rm B}$. Since dark matter remains gravitationally coupled to baryons, the Lorentz force of the PMF acting on baryons influences the clustering of the total matter component of the Universe in the post-recombination era, thereby enhancing matter power at small scales. In the post-EoR era at $z\lesssim6$, the Ly$\alpha$ and 21 cm auto- and cross-correlation spectra are biased tracers of the total matter spectrum, with the bias functions being both redshift- and scale-dependent at nonlinear scales $k\gtrsim1\:\textrm{Mpc}^{-1}$. Hence, a sub-nG PMF is expected to leave distinctive imprints on these three observables at small scales, which, as we have demonstrated in this study, may serve as important late-time probes that may be used in the near future to constrain the PMF sector. 

Our forecast analysis is based on two specific next-generation instrumental combinations, namely DESI-like+SKA1-Mid and DESI-like+PUMA, which serve as representative case studies that exploit the synergy between upcoming spectroscopic surveys and 21 cm radio facilities in probing post-EoR physics. In particular, SNR estimates for these two instrumental combinations hint that, in an idealized setting, the 21 cm auto-power spectrum may provide the highest  sensitivity to PMF-induced features. However, this apparent advantage must be interpreted with caution, as realistic foreground contamination is expected to significantly degrade the achievable SNR for 21 cm power spectrum measurements, thereby limiting their constraining power. In contrast, the Ly$\alpha$-21 cm cross-power spectrum exhibits a more robust observational outlook. Owing to its relative insensitivity to foreground contamination, it is expected to retain a substantial SNR, particularly on small scales where PMF effects are most pronounced. This positions the cross-correlation as a promising and potentially more reliable statistical probe of the PMF parameter space.
 
Further, the prospects of probing the PMF sector have been assessed through an estimation of errors in the relevant parameters by employing the Fisher forecast formalism in the light of all three post-EoR observables and both instrumental combinations. Importantly, the DESI-like+SKA1-Mid combination, which is assumed capable of probing up to $k\sim10.0\:h/\textrm{Mpc}$, is expected to provide constraints on both of the PMF parameters with an overall order-of-magnitude better level of precision (and hence with error bars that are smaller by a similar margin) than the DESI-like+PUMA combination, owing to the shorter baseline diameter of PUMA that would ideally make scales beyond $k\sim5.0\:h/\textrm{Mpc}$ inaccessible. This trend is predicted to hold true across all three of the concerned post-EoR observables. As for the observables themselves, for both instrumental combinations, a foreground-free 21 cm auto-spectrum is projected to place the tightest error bounds, being capable of furnishing relative errors $\lesssim10\%$ (and even down to $\sim1\%$) on both parameters for a sufficiently strong sub-nG PMF with a slight spectral tilt. In terms of constraining power, the foreground-free 21 cm auto-spectrum is followed by the Ly$\alpha$-21 cm cross-spectrum for DESI-like+SKA1-Mid, and by the Ly$\alpha$ auto-spectrum for DESI-like+PUMA, which may, in both cases, generally provide one order of magnitude weaker bounds on the PMF parameters than the 21 cm auto-spectrum. 

However, contamination of the 21 cm auto-spectrum by various astrophysical and terrestrial foregrounds may significantly diminish its SNR in a more realistic scenario as pointed out earlier, thereby deteriorating its ability to provide such optimistically stringent bounds on the PMF parameters. In that case, our Fisher projections concretely indicate that the uncontaminated Ly$\alpha$-21 cm cross-spectrum, probed by the DESI-like+SKA1-Mid combination, could arguably be the best channel to constrain a weakly scale-dependent sub-nG PMF based on its impact on matter clustering in the post-EoR era. From a broader scientific perspective, we highlight this prediction for the cross-correlation as one of the key findings of this work. Overall, our results underscore the importance of multi-tracer strategies, where the interplay between statistical sensitivity and systematic robustness determines the ultimate efficacy of different observables in probing primordial magnetism.

It seems prudent to ask how much improvement could be expected based on the complementarity of these late-time post-EoR observables with early-time cosmological probes. In particular, it is important to assess to what extent upcoming CMB datasets could improve the error bounds on a sub-nG PMF with a weakly scale-dependent spectrum as considered in this study. In this regard, one could reasonably expect CMB $B$-mode observations in the near future \cite{LiteBIRD:2020khw} to play an important role in constraining sub-nG PMFs \cite{Zucca:2016iur}, which could indicate the need for a combined CMB+post-EoR analysis. However, in view of the $B$-mode signal, a weakly scale-invariant non-helical PMF is known to be strongly degenerate with inflationary gravitational waves, \emph{i.e.}, it would be quite difficult to distinguish between the PMF parameters and the tensor-to-scalar ratio \cite{LiteBIRD:2024twk}, which happens to be a prominent science goal for next-generation CMB missions targeting to probe the $B$-mode signal. From this perspective, allowing room for primordial gravitational waves would make a joint forecast analysis with future CMB missions less instructive about the unique signature of such a PMF on the $B$-mode signal, which may limit its scope as far as any significant improvement in the projected bounds on the PMF parameters is concerned.  That said, the situation could be quite different for PMF spectra with sharper scale dependence, for which the aforementioned degeneracy is expected to be weaker. Consideration of such sharply scale dependent spectra falls beyond the scope of this study, and as such, we defer it to a future work.

As part of our concluding remarks, we emphasize that the present work is far from exhaustive and points towards several important directions that merit further investigation, a few of which deserve particular mention. As already highlighted, the presence of strong foregrounds in the 21 cm sector has been neglected in this study, which must be taken into account for a more realistic analysis of the 21 cm auto-spectrum and its capacity of providing constraints on the PMF sector. In addition, over the course of this study, $\Lambda$CDM-fitted nonlinear bias functions have been used to theoretically model the Ly$\alpha$ and 21 cm spectra as biased tracers of the matter spectrum at small scales, whereas a more accurate analysis demands a dedicated high-resolution MHD simulation-based computation of the bias functions for the specific PMF-driven scenario under consideration. We aim to address some of these questions in the near future.

\section*{Acknowledgments}

Authors acknowledge helpful discussions with Paulo Montero-Camacho and Debarun Paul. 
AB thanks CSIR for financial support through Senior Research Fellowship (File no. 09/0093
(13641)/2022-EMR-I). SP1 thanks CSIR for financial support through Senior Research Fellowship (File no. 09/093(0195)/2020-EMR-I). SP2 thanks ANRF, Govt. of India, for partial support through Project No. CRG/2023/003984. We acknowledge the computational facilities provided by the SyMeC HPC cluster of ISI Kolkata.

\bibliographystyle{jhep}
\bibliography{references.bib}

\end{document}